\documentclass[preprint,a4paper,5p,times,twocolumn]{elsarticle}

\usepackage{amssymb}

\usepackage{hyperref}
\newif\iflong
\longfalse

\newif\ifblind
\blindfalse

\newif\ifjournal
\journalfalse
\usepackage[inline]{enumitem}
\usepackage{ragged2e}

\usepackage{etoolbox}
\makeatletter
\patchcmd{\@makecaption}
  {\scshape}
  {}
  {}
  {}
\makeatother

\usepackage{boxedminipage}
\newenvironment{result}%
{\smallskip
	\noindent
	\let\emph=\textbf
	\begin{boxedminipage}{\columnwidth}\begin{center}\em}%
		{\end{center}\end{boxedminipage}%
	\smallskip
}

\usepackage{booktabs}
\usepackage{relsize}

\graphicspath{{./figs/}}
\DeclareGraphicsExtensions{.pdf,.png}
\usepackage{amsmath}
\usepackage{amsthm}
\usepackage{array}
\usepackage{listings}
\lstdefinelanguage{aN}[]{Python}
{
  morekeywords=[2]{int_lists},
  morekeywords=[2]{np_shapes, %
                   np_arrays, dicts, %
                   objs,
                   tuples, ints, floats, bools, lists,
                   anys, froms
                 }, %
  morekeywords=[3]{@arg,@require,@ensure},
  morekeywords=[3]{@, @generator, %
                   @exclude, @ cc_example, %
                   @timeout, @module_test}, %
  deletekeywords=[2]{min,max,int,float,list,tuple,bool,any,from,dict},
   literate={==>}{{$\Rightarrow$}}{1}
            ,
   mathescape=true,
   escapeinside={(*}{*)},
   identifierstyle=\ttfamily,
   keywordstyle=[2]{\bfseries\ttfamily},
   keywordstyle=[3]{\color{dgcol}\bfseries\ttfamily},
 }

\newcommand{\Py}[1]{\mbox{\lstinline[basicstyle=\ttfamily,language=Python]|#1|}}

\newcommand{\An}[1]{\mbox{\lstinline[basicstyle=\ttfamily,language=aN]|#1|}} %

\usepackage{multicol,multirow}
\usepackage[caption=false,font=footnotesize]{subfig}

\usepackage{tikz}
\usetikzlibrary{shapes,arrows,positioning,calc,fit,intersections,through}

\definecolor{acol}{RGB}{230,97,1}

\definecolor{lbcol}{RGB}{166,206,227}
\definecolor{dbcol}{RGB}{31,120,180}
\definecolor{lgcol}{RGB}{178,223,138}
\definecolor{dgcol}{RGB}{51,160,44}

\usepackage{xurl}
\usepackage{xspace}

\usepackage[T1]{fontenc}
\usepackage[scaled=0.81]{beramono}

\usepackage{fontawesome}
\makeatletter
\let\@afterindenttrue\@afterindentfalse
\makeatother
\usepackage{enotez}
\setenotez{list-name={URL References}}
\setenotez{counter-format=arabic}
\setenotez{mark-cs={\formatEndNoteMark}}
\newcommand{\formatEndNoteMark}[1]{\textsuperscript{\{\color{blue}{\textsf{#1}}\}}}

\DeclareDocumentCommand{\projURL}{s O{} m}
{\ifblind\endnote{\IfBooleanTF{#1}{{\color{blue}Omitted for double-blind review}}{\IfNoValueF{#2}{#2\xspace}\url{#3}}}\else\endnote{\IfNoValueF{#2}{#2\xspace}\url{#3}}\fi}

\newcommand{\annotest}{{\smaller[0.5]{\textsc{aNNo\-Test}}}\xspace}
\newcommand{\an}{{\smaller[0.5]{\textsc{aN}}}\xspace}

\newcommand{\nicepar}[1]{\textbf{#1}}

\newif\ifdraft
\draftfalse

\ifdraft
\newcommand{\note}[1]{\noindent\textcolor{blue}{NOTE. #1}}
\newcommand{\caf}[1]{\noindent\textcolor{teal}{CAF: #1}}
\newcommand{\mr}[1]{\noindent\textcolor{blue}{MR: #1}}
\newcommand{\moe}[1]{\textcolor{blue}{Moe: #1}}
\newcommand{\question}[1]{\noindent\textcolor{red}{QUESTION. #1}}
\newcommand{\copied}[1]{\noindent\textcolor{magenta}{COPIED. "#1"}}
\else
\newcommand{\note}[1]{}
\newcommand{\caf}[1]{}
\newcommand{\mr}[1]{}
\newcommand{\moe}[1]{}
\newcommand{\question}[1]{}
\newcommand{\copied}[1]{}
\fi

\usepackage{adjustbox}

\newcolumntype{R}[2]{%
    >{\adjustbox{angle=#1,lap=\width-(#2)}\bgroup}%
    l%
    <{\egroup}%
}

\ifjournal
  \journal{Journal of Systems and Software}
\else
  \journal{}
\fi

\usepackage[switch]{lineno}

\begin{document}

\begin{frontmatter}

\title{\textsf{An Annotation-based Approach for Finding Bugs in Neural Network Programs}}
\date{}

\ifjournal
\else
\author{Mohammad Rezaalipour}
\author{Carlo A. Furia}

\affiliation{organization={Software Institute, USI Università della Svizzera italiana},%
            city={Lugano},
            country={Switzerland}}
\fi

\begin{abstract}
As neural networks are increasingly included as
core components of safety-critical systems,
developing effective testing techniques
specialized for them becomes crucial.
The bulk of the research 
has focused on testing neural-network \emph{models};
but these models are defined by writing programs, 
and there is growing evidence that these 
\emph{neural-network programs} often have bugs too.
This paper presents \annotest: 
an approach to generating test inputs for neural-network 
programs.
A fundamental challenge is that
the dynamically-typed languages (e.g., Python)
commonly used to program neural networks
cannot express 
detailed constraints about valid function inputs
(e.g., matrices with certain dimensions).
Without knowing these constraints, automated test-case generation
is prone to producing invalid inputs, which
trigger spurious failures and are useless
for identifying real bugs.
To address this problem, we introduce a simple
annotation language tailored for concisely expressing
valid function inputs in neural-network programs.
\annotest takes as input an annotated program,
and uses property-based testing to
generate random inputs that satisfy the validity constraints.
In the paper, 
we also outline guidelines that 
simplify writing \annotest annotations.

We evaluated \annotest on 19 neural-network programs
from Islam et al's survey.~\cite{Islam:2019},
which we manually annotated following our guidelines---%
producing 6 annotations per tested function on average.
\annotest automatically generated test inputs
that revealed 94 bugs, including 63 bugs that 
the survey reported for these projects.
These results suggest that \annotest 
can be a valuable approach to finding
widespread bugs in real-world neural-network programs.
\end{abstract}

\end{frontmatter}

\section{Introduction}
\label{sec:introduction}

Neural networks have taken the (programming) world by storm.
With their capabilities of solving tasks 
that remain challenging for traditional software,
they have become central components of 
software systems implementing complex functionality 
such as image processing, speech recognition, and natural language processing,
where they can reach performance at or near human level.
These tasks are widely applicable to domains
such as automotive and healthcare,
where safety, reliability, and correctness are critical.
Therefore, the software engineering (research) community has been hard at work
designing techniques to assess and ensure the dependability of software
with neural network (NN) components.

Testing techniques, in particular, are being extensively developed
to cater to the specific requirements of NN (and, more generally,
machine learning) systems~\cite{MLTesting-survey}.
Most of this research focuses on testing NN \emph{models}:
instances of a specific NN architecture,
trained on some data and then used
to classify or transform new data.
Testing a NN model entails assessing qualities such as
its robustness and performance as a classifier.
However, neural networks are programs too:
a NN model is usually implemented
in a programming language like Python, 
using frameworks such as Keras or TensorFlow.
There is clear evidence that these \emph{neural network programs}
tend to be buggy~\cite{Islam:2019,Humbatova:2019};
therefore, a technique for finding these bugs 
would be practically very useful and complement 
the extensive work on NN model testing~\cite{Zhang:2022}.
This paper presents a novel contribution in this direction.

NN programs may seem simple by traditional metrics of complexity:
for example, the average project size of the NN projects surveyed by Islam et al.~\cite{Islam:2019} is just 2165
lines of code; and the majority of
the bugs they found are relatively simple ones such as crashes and API misuses.
Nevertheless, other characteristics
make traditional test-case generation techniques
ineffective to test such programs.
NN programs are written in dynamically typed
languages like Python, where the type of variables is unknown statically.
Without this information, generating valid inputs is challenging for generic
techniques such as random testing and genetic algorithms~\cite{Lukasczyk:2020}.
Even if type annotations %
were available,
NN programs routinely manipulate complex data structures---%
such as vectors, tensors, and other objects---%
whose precise ``shape'' is not expressible with the standard types
(integers, strings, and so on).
As we demonstrate in \autoref{sec:example} and \autoref{sec:rq4-answer},
without such precise information automated test case generation
tends to generate many invalid inputs that trigger spurious failures.

This paper presents \annotest: an approach to automatically generating
bug-finding inputs for NN program testing.
A key component of \annotest (described in \autoref{sec:annotest})
is \an: a simple annotation language
to concisely and precisely express the valid inputs
of functions in NN programs.
The \an language supports expressing the kinds of constraints
that are needed in NN programs
(for example: a variable should be a vector of size from 2 to 5 with components that are positive integers).
\an is also easily extensible to
accommodate other constraints that a specific NN program may need to encode.

Given an annotated NN program, \annotest automatically generates 
unit tests for the program that span the range of valid inputs.
To this end, the current implementation of \annotest
uses property-based testing (more precisely,
the Hypothesis~\cite{Hypothesis} test-case generator).
Using the \an language decouples
specifying the constraints from the back-end used to generate
the actual tests;
therefore, different back-end tools could also be used
that better suite the kinds of constraints used in a project's annotations.

\autoref{sec:experiments} describes an extensive experimental evaluation
of \annotest, targeting 19 open-source NN programs, manually analyzed by Islam et al.~\cite{Islam:2019},
using some of the most widely used NN frameworks (Keras, TensorFlow, and PyTorch).
After we manually annotated 24 functions, \annotest generated tests
triggering 63 known bugs reported by Islam et al.~\cite{Islam:2019} for these functions,
as well 31 previously unknown bugs.
To experiment with \annotest's capabilities when used extensively,
we also annotated all functions in two larger NN projects;
the total of 330 annotations that we wrote enabled \annotest to discover
50 
bugs
with only 
6
false positives.
These experiments demonstrate that \annotest can be used both extensively
on a whole project, and opportunistically on only a few selected functions
that are critical.
Since our evaluation is based on Islam et al.~\cite{Islam:2019}'s extensive survey, %
it can assess \annotest's capabilities of finding relevant bugs in real-world NN programs.
In other experiments, we quantify the amount of annotations needed by \annotest,
compare it to generic (non NN-specific) test-case generators for Python,
as well as to developer-written tests,
so as to better understand the trade-off between
programmer effort and quality assurance benefits
it offers.

In summary, this paper makes the following contributions:
\begin{itemize}
\item \annotest: an approach for test-case generation
  geared to the characteristics of NN programs.
\item \an: a simple annotation language capable of
  concisely expressing precise constraints on the valid inputs of functions in NN programs,
  with basic guidelines to use it.
\item An experimental evaluation of \annotest's bug-finding capabilities on
  19 open-source NN projects surveyed by Islam et al.~\cite{Islam:2019}.
\item For reproducibility, the implementation of \annotest and all experimental
  artifacts are publicly available:
  \begin{center}
    \url{https://figshare.com/s/00ef658a6a51cccbaed6}
  \end{center}
\end{itemize}

In the paper, we refer to several URLs in order to document specific parts of a project's source code.
We introduce these references by means of superscript numeric marks,
and list them at the end of the paper after the usual bibliographic references.
These superscripts are in blue between curly braces\projURL[An example of URL reference: ]{https://bugcounting.net}
so that they can be easily distinguished from regular footnotes.

\nicepar{Scope.}
While \annotest is applicable, in principle, to any Python programs---not just NN programs---
it was designed to primarily cater to the characteristics of NN programs.
As we will see concretely with \autoref{sec:example}'s example,
NN programs often involve complex constraints on their inputs,
which are impossible or highly impractical to express using Python's type hints annotations.
\annotest provides annotations that go beyond type hints,
and hence are especially useful for the kinds of constraints that we commonly find in NN programs.
On the other hand,
being able to \emph{express} complex constraints is not sufficient to build tests automatically;
as we will see in~\autoref{sec:rq4-answer},
\emph{generating} inputs that satisfy the constraints is challenging;
simple strategies such as generating input at random and then filtering them using the constraints
are mostly ineffective.
\annotest defines suitable generators for each of its constraints,
so that valid inputs can be generated efficiently and automatically
even for the complex combinations of input constraints
that are common in NN programs.

\section{An Example of Using \annotest}
\label{sec:example}

DenseNet\projURL[DenseNet project page:]{https://github.com/cmasch/densenet/}
is a small Python library that implements
densely connected convolutional networks~\cite{Huang:2019}
(a NN architecture
where each layer is directly connected to every other layer)
on top of the Keras framework.
\autoref{lst:motivational1} shows a slightly simplified
excerpt of function \Py{DenseNet}---%
the main entry point to the library---%
in an earlier version of the
project.\projURL{https://github.com/cmasch/densenet/blob/70ee31d0f6f800324fbe98ea687122395248d39e/densenet.py}

\begin{lstlisting}[float=*,caption=An excerpt of function \Py{DenseNet} from project DenseNet. The code has a bug on line~\ref{l:bug1}., label={lst:motivational1}]
def DenseNet(input_shape=None, dense_blocks=3, dense_layers=-1, growth_rate=12, nb_classes=None,
             dropout_rate=None, bottleneck=False, compression=1.0, weight_decay=1e-4, depth=40):
   if nb_classes == None:
      raise Exception('Please define number of classes')
   if compression <= 0.0 or compression > 1.0:
      raise Exception('Compression must be between 0.0 and 1.0.')
   if type(dense_layers) is list:
      if len(dense_layers) != dense_blocks:
         raise AssertionError('Dense blocks must be the same as layers')(*\label{l:exception}*)
   elif dense_layers == -1:
      dense_layers = (depth - 4) / 3    (*\textcolor{red}{\# Bug: division / returns a float}*) (*\label{l:bug1}*)
   # ... 23 more lines of code ...
\end{lstlisting}

\begin{lstlisting}[float=*,caption=\an annotations for function \Py{DenseNet} in \autoref{lst:motivational1}., label={lst:motivational1_annotation},language=aN]
@arg(input_shape): tuples(ints(min=20, max=70), ints(min=20, max=70), ints(min=1, max=3)) (*\label{l:an:input_shape}*)
@arg(dense_blocks): ints(min=2, max=5) (*\label{l:an:dense_blocks}*)
@arg(dense_layers): anys(-1, ints(min=1, max=5), int_lists(min_len=2, max_len=5, min=2, max=5)) (*\label{l:an:dense_layers}*)
@arg(growth_rate): ints(min=1, max=20) (*\label{l:an:growth_rate}*)
@arg(nb_classes): ints(min=2, max=22) (*\label{l:an:nb_classes}*)
@arg(dropout_rate): floats(min=0, max=1, exclude_min=True, exclude_max=True) (*\label{l:an:dropout_rate}*)
@arg(bottleneck): bools() (*\label{l:an:bottleneck}*)
@arg(compression): floats(min=0, max=1, exclude_min=True) (*\label{l:an:compression}*)
@arg(weight_decay): floats(min=1e-4, max=1e-2) (*\label{l:an:weight_decay}*)
@arg(depth): ints(min=10, max=100) (*\label{l:an:depth}*)
@require(type(dense_layers) is not list or len(dense_layers)==dense_blocks)(*\label{l:an:require}*)
\end{lstlisting}

The complete implementation of function \Py{DenseNet} comprises
34 lines of code (excluding comments and empty lines),
and follows a straightforward logic:
after checking the input arguments %
(code in~\autoref{lst:motivational1}),
it combines suitable instances of Keras classes to model
a densely connected network, and finally returns a model object to the caller.
\autoref{lst:motivational1}'s code, however, has a 
bug at line~\ref{l:bug1}---%
one of the bugs collected in Islam et al.~\cite{Islam:2019}'s survey.
The expression assigned to \Py{dense_layers}
is a floating point number because
the division operator \Py{/} always returns a float in Python~3;
however, if \Py{dense_layer} is not an integer, 
a later call in \Py{DenseNet}'s code to the Keras library fails.
DenseNet's developers discovered the bug and fixed it
(by adding an \Py{int} conversion at line~\ref{l:bug1})
in a later project revision.\projURL{https://github.com/cmasch/densenet/commit/693d772ae9dcdb4d524b25d7d2f6428de4a524ff\#diff-813086a9be01b05b352f0111384c48e74735b009e22f4bab1f3dcaa06e2303c2R68}

\Py{DenseNet}'s implementation is deceptively simple:
despite its small size and linear structure,
it only accepts input arguments in very specific ranges.
Argument \Py{input_shape}, for example, corresponds to a so-called
\emph{shape tuple} of integers;
in \Py{DenseNet}, it should be a triple of integers with first element at least 20.
If the first element %
is less than 20,
\Py{DenseNet} eventually fails while trying to create a layer
with a negative dimension---which violates an assertion of the Keras library.
Another example is argument \Py{dense_layers}, which
can be an integer or an integer list;
if it is the latter, its length must be equal to argument \Py{dense_block},
or \Py{DenseNet} terminates at line~\ref{l:exception} with an assertion violation.

Without knowing all these details about valid inputs,
testing \Py{DenseNet} using a general-purpose automated test-case generator
would trigger lots of spurious failures\footnote{
  For example, Pynguin~\cite{Lukasczyk:2020}
  generates 8 tests, all invalid and none triggering the failure at line~\ref{l:bug1}.
  With type hints (supported by Pynguin),
  it generates 5 tests, 4 invalid and none triggering (any) failure.
  \autoref{sec:rq4-answer} describes more experiments with Pynguin.
  (As we discuss in \autoref{sec:experimental_subjects},
  Pynguin doesn't work with the version of TensorFlow used by \autoref{lst:motivational1}'s code;
  thus, we mocked the relevant library calls in this example.)
}
when executing tests that call \Py{DenseNet} with invalid inputs.
The few failing but valid tests that trigger bugs such as that in \autoref{lst:motivational1}
would be a needle in the haystack of all invalid tests,
thus essentially making automated test-case generation of little help
to speed up the search for bugs.

To precisely and concisely express the complex constraints
on valid inputs that often arise in NN programs,
we designed the \an annotation language---which is a central component
of the \annotest approach.
\autoref{lst:motivational1_annotation} shows annotations
written in \an\footnote{The \an annotations in the paper use a pretty-printed and slightly simplified syntax.} that characterize \Py{DenseNet}'s valid inputs.
Whereas \autoref{sec:annotest} will present \an's features in greater detail,
it should not be hard to glean the meaning of the annotations in \autoref{lst:motivational1_annotation}.
For example, the first annotation encodes the aforementioned constraint on \Py{input_shape}, and the last one expresses the relation between \Py{dense_layers} and \Py{dense_blocks} when the former is a list.
It should also be clear that \an's expressiveness is much greater than
what is allowed by the standard programming-language types---%
such as Python's type hints.\projURL[Type hints: ]{https://www.python.org/dev/peps/pep-0484/}

Equipped with the annotations in \autoref{lst:motivational1_annotation},
\annotest generates and runs 36 unit tests for \Py{DenseNet} in 53 seconds.
All the tests are valid, and only one is failing,
reaching \autoref{lst:motivational1}'s line~\ref{l:bug1}
and then ending with a failure due to \Py{dense_layers} being a float
that we described above---precisely revealing the bug.

The experimental evaluation of \annotest---%
described in \autoref{sec:experiments}---%
will analyze many more NN programs whose characteristics,
input constraints, and faulty behavior are along the same lines as
the example discussed in this section.
This will demonstrate \annotest's
capabilities of precisely testing and finding bugs in NN programs.

\section{How \annotest works}
\label{sec:annotest}

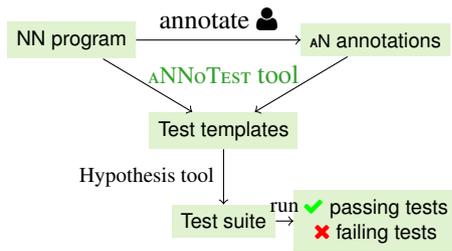
\begin{figure}[tb]
    \centering
    \begin{tikzpicture}[
    stage/.style={rectangle,fill=lgcol!40!white,
                font=\footnotesize\sffamily,text=black, draw=none},
    tool/.style={draw=none,fill=none,font=\footnotesize},
    node distance=12mm and 20mm, on grid,auto,
    align=center
    ]
    
    \node[stage] (src) {NN program};
    \node[right=of src] (mid) {};
    \node[stage,right=of mid] (annotations) {\an annotations};
    \node[stage,below=of mid] (hypo) {Test templates};
    \node[stage,below=of hypo] (tests) {Test suite};
    \node[stage,right=of tests] (pass-fail) {{\color{green} \faCheck} passing tests\\{\color{red} \faClose} failing tests};
    
      \begin{scope}[->]
      \draw (src) to node[above] {annotate \faUser} (annotations);
      \draw (src) to (hypo);
      \draw (annotations) to (hypo);
      \node[below=4.5mm of mid] {\color{dgcol} \annotest tool};
      \draw (hypo) to node[left] {\footnotesize Hypothesis tool} (tests);
      \draw (tests) to node[above] {\footnotesize run} (pass-fail);
      \end{scope}
    \end{tikzpicture}

    \caption{An overview of how the \annotest approach works.}
    \label{fig:annotest_phases}
    \vspace{2mm}
  \end{figure}

\autoref{fig:annotest_phases}
overviews the overall process
followed by the \annotest approach.
To test a NN program with \annotest,
we first have to annotate its functions
(including member functions, that is methods) %
using the \an annotation language
(\autoref{sec:an_language}).
This is the only step that is manual,
since the annotations
have to encode valid inputs of the tested functions%
---the same kind of information that is needed to write unit tests.
\autoref{sec:guidelines}
provides guidelines that help structure the manual
annotation process so that it only requires a
reasonable amount of effort;
furthermore, users do not need to annotate a whole program
but only those functions that they want to test with \annotest.
Then, the \annotest tool
takes as input an annotated program and
generates unit tests for it.
To this end, it encodes the constraints
expressed by the \an annotations
in the form of test templates for
the property-based test-case generator
Hypothesis
(\autoref{sec:test_generation});
then, it runs Hypothesis which takes care of generating suitable tests.
Finally, the generated unit tests can be run as usual
to find which are passing and which are failing---%
and thus expose some bugs in the NN program (\autoref{sec:oracle_problem}).

\subsection{The \an Annotation Language}
\label{sec:an_language}

By writing annotations in the \an language,
developers can precisely express the valid inputs of a function
in a NN program.\footnote{Directly annotating the source code, rather than having a separate generator used only when testing, also helps keep the annotations consistent with the implementation.}
To this end, \an provides \emph{type} annotations
(\autoref{sec:annotations-argument})
and \emph{preconditions}
(\autoref{sec:annotations-preconditions}),
as well as an extension mechanism to define arbitrarily complex constraints
(\autoref{sec:annotations-generators}).
In addition, 
\an offers a few \emph{auxiliary} annotations (\autoref{sec:annotations-auxiliary}),
which encode other kinds of information that is practically
useful for test-case generation.

\subsubsection{Type Annotations}
\label{sec:annotations-argument}

A \emph{type} annotation follows the syntax \An{@arg(v): $T$},
where \An{v} is a function argument (parameter), and $T$ is
a \emph{type constraint} that specifies a set of possible values for \An{v}.
A type annotation refers to the function that immediately follows
it in the source code.
A function can have up to as many type annotations as it has arguments.

\an supports several different \emph{type constraints},
which can express a broad range of constraints---from
simple ones, such as those that are also expressible using
Python's type hints, up to complex instances of special-purpose classes.
The simplest, and most specific, type constraint
uses keyword \An{froms}\footnote{\an type constraints use names that
  are ``pseudo-plurals'' (by adding a trailing \Py{s})
  of the corresponding Python types. This avoids using reserved keywords and also conveys the idea that a type constraint identifies a set of values.
This convention is also customary in property-based testing~\cite{QuickCheck}.}
to enumerate a list of valid values.
For example, constraint
\An{froms([0, 0.0, None, zero()])} corresponds to any of the four values:
integer zero, floating-point zero, \Py{None},
and what is returned by the call \Py{zero()}.

Constraints for \nicepar{atomic types}
specify that an argument is a Boolean (\An{bools}),
an integer number (\An{ints}),  or a floating-point number (\An{floats}).
Integer arguments can be restricted to a range between \An{min}
and \An{max} values; for example, \autoref{lst:motivational1_annotation}'s
line~\ref{l:an:nb_classes}
constrains \Py{nb_classes} to be an integer between 2 and 22.
Floating-point arguments can also be restricted to ranges,
and the ranges can be open, closed, or half-open;
for example, \autoref{lst:motivational1_annotation}'s
line~\ref{l:an:compression}
constrains \Py{compression} to be a number in the half-open interval $(0, 1]$
which includes 1 but excludes 0.
Floating-point constraints also support including or excluding
the special values \Py{NaN} and \Py{Inf}, as well as the precision (in bits)
of the generated floating point values.

Constraints for \nicepar{sequences}
specify that arguments are Python \An{lists}, \An{tuples}, or
an array in the NumPy\projURL[NumPy: ]{https://numpy.org/} library
(which is widely used in NN programs, as well as other data-intensive applications).
Lists and tuples can have any number of elements,
whose possible values are also constrained using \an's type constraints.
For example, \autoref{lst:motivational1_annotation}'s line~\ref{l:an:input_shape}
specifies a tuple with 3 integer elements: the first and second one between
20 and 70, and the third one between 1 and 3.
\an also includes shorthands for lists with homogeneous elements:
\autoref{lst:motivational1_annotation}'s line~\ref{l:an:dense_layers}
uses shorthand \An{int_lists} to specify lists of length between 2 and 5, whose elements are integers between 2 and 5.

The \emph{shape} of a NumPy array is a tuple of positive integers
that characterize its size.
For example, the tuple \Py{(256, 256, 3)} is the shape of a 3-dimensional
array whose first two dimensions have size 256 and whose last dimension has size 3;
arrays with this shape can represent 256x256 pixel color pictures.
Type constraint \An{np_shapes} specifies arguments that represent shapes
with a certain range of possible dimensions and sizes.
For example, \An{np_shapes(min_dims=3, max_dims=3)}
are the shapes of all 3-di\-men\-sion\-al arrays whose dimensions can have any size.

Type constraint \An{np_arrays}
specifies NumPy array arguments with any shape
and whose elements have any of the valid NumPy types.
The shape can be constrained by an \An{np_shapes} annotation
or given directly as a tuple.
For example,
using the shape mentioned in the previous paragraph,
$\An{np_arrays(np_type=dtype("uint32")}, \linebreak \An{shape=(256,256,3))}$
specifies 256x256x3 arrays whose components are unsigned 32-bit integers (one of NumPy's data-types),
which could represent random color pictures.

Type constraints for \nicepar{maps}
specify Python's widely used associative dictionaries:
\An{dicts($K$, $V$, min_size, max_size)}
corresponds to all subsets
of the Cartesian product $K \times V$
with between \An{min_size} and \An{max_size} elements,
where $K$ and $V$ are type constraints
that apply to the keys and values respectively.
A typical usage of this is to constraint
Python's optional keyword argument \Py{**kwargs}.
For example,
\autoref{lst:example_keyword_arguments}
shows how we used \An{dicts}
to constrain the \Py{**kwargs} 
argument of function \Py{dim_ordering_reshape}\projURL{https://github.com/bstriner/keras-adversarial/blob/master/examples/image_utils.py#L34}
(from a project using NN models to simulate multi-player games),
so that it simply consists
of all mappings from string 
\Py{"input_shape"}
to singletons representing the shapes of monodimensional arrays.

\begin{lstlisting}[float=*,caption={An example of \an annotations for a function with keyword arguments.}, label={lst:example_keyword_arguments},language=aN]
@arg(k): ints(min=1, max=1000)
@arg(w): ints(min=1, max=1000)
@arg(kwargs): dicts(keys=froms(["input_shape"]), values=np_shapes(min_dims=1, max_dims=1)) (*\label{l:an:kwargs}*)
def dim_ordering_reshape(k, w, **kwargs):
\end{lstlisting}

To express the \nicepar{unions} of several type constraints,
\an includes the \An{anys} type constraint,
which specifies the union of its arguments.
For example,
\autoref{lst:motivational1_annotation}'s line~\ref{l:an:dense_layers}
says that \Py{dense_layers} can be any of:
\begin{enumerate*}[label=\emph{(\roman*)}]
\item the number $-1$,
\item an integer between $1$ and $5$, or
\item an integer list with between $2$ and $5$ elements that are between $2$ and $5$.
\end{enumerate*}

\subsubsection{Custom Generators}
\label{sec:annotations-generators}

While \an's type annotations can define a broad range of frequently used
constraints, they cannot cover all cases that one may encounter in practice.
To support \emph{arbitrary} type constraints,
\an includes the \An{objs(gen)} annotation.
This is used as a type constraint,
and identifies all values that are produced by the user-provided
\emph{generator} function \Py{gen}.
Function \Py{gen} must be visible at the entry of the functions whose
annotations refer to it;
\Py{gen} itself is marked with the annotation \An{@generator}.

\begin{lstlisting}[float=bt,caption={An example of using type constraint \An{objs} and a custom generator function.}, label={lst:example_objs},language=aN,xrightmargin=-1mm]
@arg(generator): objs(gan_gens)
@arg(discriminator): objs(gan_discs)
@arg(name): froms(["gan1", "gan2", "gan3",
                   "gan4", "gan5"])
def build_gan(generator, discriminator,
              name="gan"):
   # ... 

@generator
@exclude (*\label{l:gan:exclude}*)
@arg(latent_dim): ints(min_value=1, 
                       max_value=1000)
@arg(input_shape): np_shapes(min_dims=2)
def gan_gens(latent_dim, input_shape):
    from examples.example_gan import model_generator
    generator = model_generator(latent_dim,
                                input_shape)
    return generator
\end{lstlisting}

For instance, \autoref{lst:example_objs}
shows the annotations
we wrote for function \An{build_gan}\projURL{https://github.com/bstriner/keras-adversarial/blob/master/keras_adversarial/adversarial_utils.py#L10}
(from the same project as \autoref{lst:example_keyword_arguments}).
The function combines 
two Keras model
instances,
\Py{generator}\footnote{It is just a coincidence that one argument is also named ``generator''.}
and \Py{discriminator}, 
to build GANs (Generative Adversarial Networks~\cite{Goodfellow:2014}).
These instances are complex objects that
are built by calls to the Keras library;
therefore, we introduced two custom generators, \Py{gan_gens} and \Py{gan_discs}, 
that construct such instances for testing \Py{build_gan}.
\autoref{lst:example_objs}
shows \Py{gan_gens}'s implementation:
the generator's input are constrained by using \an's type annotations as usual;
\annotest will use \Py{gan_gens}'s output
as input for \Py{build_gan}.

Whereas generators such as \Py{gan_gens} may look daunting to write at first,
we found that they
simply encapsulate existing snippets of the project that
call the function under test (\Py{build_gan} in \autoref{lst:example_objs}).
Based on this observation,
\autoref{sec:generator-functions}
presents a simple process to build
generators by combining common refactoring steps;
this drastically alleviates the effort
to write generators, reducing it to just selecting the right snippets of client code in the project.

\subsubsection{Preconditions}
\label{sec:annotations-preconditions}

Argument annotations constrain each function argument individually.
\emph{Preconditions} may express constraints that affect multiple arguments simultaneously:
\An{@require($P$)}, where $P$ is a Python Boolean expression,
specifies that a function's arguments must be such that $P$ evaluates to true.
A precondition refers to the function that
immediately follows it in the source code.
Expression $P$ may refer to any
arguments of the specified function,
as well as to any other program element
that is visible at the function's entry
(such as other class members).
A function can have any number of preconditions,
all of which constraint the function's argument.
For example,
\autoref{lst:motivational1_annotation}'s line~\ref{l:an:require}
requires that, whenever argument \Py{dense_layer} is a list,
it should have
as many elements as the value of integer argument \Py{dense_blocks}.

\subsubsection{Auxiliary Annotations}
\label{sec:annotations-auxiliary}

The \an language includes a few more features to control
the test-generation process.
Functions marked with \An{@exclude} are not tested
(such as generator \Py{gan_gens} in \autoref{lst:example_objs}).
Annotation \An{@timeout} introduces a timeout to the unit tests
generated for the function it refers to.

Python modules may include snippets of code that is not
inside any functions or methods but belongs to an
implicit ``main'' environment.
\annotest will generate tests for this environment
for any module that is annotated with \An{@module_test}.
Since modules don't have arguments, these tests simply import and execute
the main environment.
This is a simple feature, but practically useful 
since some of the NN program bugs that were surveyed~\cite{Islam:2019}
are located in the main environment.

\begin{lstlisting}[float=bt,caption={An example of using the \An{cc_example} auxiliary annotation on the constructor of class \Py{ImageGridCallback}.}, label={lst:example_cc_example},language=aN,xrightmargin=0mm]
@arg(image_path):
   froms(["image1.png", "image2.png",
          "image3.png", "image4.png"])
@arg(generator): objs(grids)
@arg(cmap):
  froms(['gray', 'bone', 'pink',
         'spring', 'summer', 'cool'])
@cc_example(["image1.png",
             grids(3, 6, 6, 3), 'gray'])
def __init__(self, image_path,
             generator, cmap='gray'):
  # ... 
\end{lstlisting}

To test an instance method \Py{m}, 
one needs to generate an instance \Py{o}
of \Py{m}'s class \Py{C} to use as target of the call to \Py{m}.
To this end, \Py{C}'s constructor is called. %
The constructor may also be equipped with \an annotations;
as a result, testing \Py{m} entails also testing \Py{C}'s constructor.
This can be a problem if the constructor has bugs that prevent
a correct execution of \Py{m}.
To handle this scenario, \an includes the annotation \An{@cc_example},
which supplies a constructor with a list of concrete
inputs for it.
If \Py{C}'s constructor is equipped with this annotation,
\annotest will only call it using the inputs given by the \An{@cc_example}
annotation when it needs to create instances to test any methods of \Py{C}.
This way, one can effectively decouple testing a class's constructor
from testing the class's (regular) methods, so that any bugs in the former
do not prevent testing of the latter.
For example,
the constructor of class \Py{ImageGridCallback}\projURL{https://github.com/bstriner/keras-adversarial/blob/master/keras_adversarial/image_grid_callback.py#L7}
shown in \autoref{lst:example_cc_example}
is regularly tested through its type annotations;
however, when it is used to construct instances of the class
to test other methods,
it is only called with the more restricted set of inputs
specified by the \An{@cc_example} annotation.
The example also demonstrates that a generator function
(\Py{grids} in this case)
can also be used as a regular function
(second component of \An{@cc_example}).

\subsection{Annotation Guidelines}
\label{sec:guidelines}

To test a NN program using \annotest, one must first annotate the functions to be tested using the language described in~\autoref{sec:an_language}.
Ultimately, writing suitable annotations requires knowledge about the program's
specification---that is, its intended behavior.
The very same knowledge is necessary to write \emph{unit tests}
for the programs; the only difference is that
a test supplies individual (valid) inputs, whereas an annotation
can capture a range of possible (valid) inputs.

This entails that the effort of writing annotations (or tests)
for a project depends on whether the programmer already has this knowledge%
---typically, because they are developers of the project under test---%
or is trying to test a project they are not familiar with.
In this section, we focus on the latter, more challenging scenario.
To help such a process of ``discovery''---%
figuring out suitable annotations for NN programs written by others---%
and to make it cost-effective, 
we present some simple guidelines
that suggest which artifacts to inspect and in which order.
In the experiments described in \autoref{sec:experiments},
we followed these guidelines to annotate NN projects
systematically and with reasonable effort---despite our previous lack of familiarity with those codebases.

\begin{table}[!htb]
  \setlength{\tabcolsep}{2pt}
  \begin{tabular}{r p{0.47\columnwidth} p{0.47\columnwidth}}
    \toprule
    & \textsc{source} & \textsc{annotations} \\
    \midrule
    1 & calls of \Py{f} in its project \Py{P}
                    &
                      basic type annotations \An{@arg}
    \\
    2 & assertions and exceptions raised by \Py{f}'s implementation
                    &
                     refined type annotations \An{@arg},
                      preconditions \An{@require}
    \\
    3 & calls of NN framework functions in \Py{f}'s implementation
                    &
                      refined type annotations \An{@arg},
                      preconditions \An{@require}, \newline
                      custom generators
    \\
    4 & calls of other functions \Py{g} in \Py{P} %
                    &
                      annotations of \Py{g}
    \\
    \bottomrule
  \end{tabular}
  \caption{Guidelines to inspect the implementation of a NN function \Py{f} to
  suggest how to annotate it using \an's annotation language. Each \textsc{source} of information in \Py{f} or elsewhere in \Py{f}'s project \Py{P} suggests matching \an \textsc{annotations}.}
  \label{tab:guidelines}
\end{table}

Consider a Python function \Py{f} in
some NN project \Py{P} that we would like to test.
If \Py{f}'s behavior (and, in particular, the constraints on its inputs)
is documented in the project, this documentation should be the first source
of information to write \an annotations.
However, if \Py{f} lacks any (precise) documentation,%
\footnote{Many of the NN programs we used in \autoref{sec:experiments}'s experiments are sparsely documented.}
we will have to inspect its implementation.
\autoref{tab:guidelines} lists four sources of information about \Py{f}'s valid inputs in increasing level of detail.

To bootstrap the process, we inspect any usage of \Py{f} within the NN program \Py{P}.
Since we focus on testing \emph{programs}, not libraries,
it's likely that every major function is called somewhere in \Py{P}.
These calls of \Py{f} provide basic examples of valid inputs,
which we loosely encode using \an's type annotations of \autoref{sec:annotations-argument}.
In \autoref{lst:motivational1}'s example,
looking at usages of \Py{DenseNet} indicates %
that \Py{input_shape} should be a triple
of \Py{int},
\Py{compression} should be a \Py{float},
and so on.%
\footnote{For example, the \Py{README.md} file in DenseNet's repository
  presents an example of using function \Py{DenseNet} where argument \Py{input_shape} is set to the triple
  \Py{(28, 28, 1)}.}

Next, we look into \Py{f}'s implementation
for any (implicit or explicit) \emph{input validation}.
Often, a function uses exceptions or assertions
to signal invalid input arguments.
This information is useful to \emph{refine}
the basic type annotations,
and may also suggest constraint that involve multiple arguments---%
which we can encode using \an's preconditions of \autoref{sec:annotations-preconditions}.
In \autoref{lst:motivational1}'s example,
\Py{DenseNet}'s initial validation
clearly
indicates, among other things, \Py{compression}'s precise interval of validity,
and the precondition on line~\autoref{l:an:require} in \autoref{lst:motivational1_annotation}.

The library functions from some NN framework
used in \Py{f}'s implementation may also (indirectly)
introduce requirements on \Py{f}'s inputs
or otherwise suggest plausible ranges of variability.
Indirect constraints may be more complex, %
and may even require custom generators (\autoref{sec:generator-functions}).
In the running example,
a call to Keras's \Py{Convolution2D} constructor
in \Py{DenseNet}
(not in \autoref{lst:motivational1})
suggests the range for argument \Py{weight_decay}
at line~\autoref{l:an:weight_decay} in \autoref{lst:motivational1_annotation}.

Whenever \Py{f}'s implementation calls
other functions in the same project, this process can be
repeated for these other functions,
thus ensuring the consistency of the other functions' and \Py{f}'s annotations.
In the running example,
\Py{DenseNet}
calls in a loop another function \Py{dense_block} in the same project,
passing \Py{growth_rate} as argument and then incrementing it in
each iteration.
The input constraints of \Py{dense_block}, once figured out,
indirectly suggest the validity range for \Py{DenseNet}'s \Py{growth_rate}
at line~\autoref{l:an:growth_rate} in \autoref{lst:motivational1_annotation}.

The guidelines we described are flexible and remain useful even if
they are not followed in full.  For example, sometimes we found it
useful to start from very narrow annotations (merely encoding the
available examples of usages of \Py{f} in \Py{P}) and relax them
as we discovered more information---rather than going from basic
to specific as we did in most examples---since this allowed us to
generate some sample tests early on.
The guidelines are also applicable with different levels of exhaustiveness,
regardless of whether your goal is to annotate as much as possible
in a project, or just test a few selected functions.
In the former case, it is advisable to start annotating the simplest, shortest
functions, so that their annotations can then suggest how to annotate
the more complex, longer ones.

\subsection{Building Custom Generators by Refactoring}
\label{sec:generator-functions}

As presented in \autoref{sec:annotations-generators},
annotation \An{@arg(a): @objs(f)}
tells \annotest to use a \emph{custom generator} function \An{f}
in order to build suitable inputs for some argument \An{a}.
In principle, \An{f} may be an arbitrarily complex piece of code;
in practice, we found that the very projects
we are annotating already include snippets of code that
can be reused as generators of complex objects.
In this section, we demonstrate, on an example, how to build such generators
by applying a few refactoring operations to the relevant snippets of code.
Modern IDEs such as PyCharm\projURL[The PyCharm Python IDE]{https://www.jetbrains.com/pycharm/}
can automate such refactoring steps.
This drastically reduces the effort of
building custom generators to just selecting the right snippets of code and doing some copy-pasting in the IDE.

\autoref{lst:generator-example-part1}
shows the signature of
function \Py{G_convblock}\projURL{https://github.com/naykun/TF_PG_GANS/blob/master/Tensorflow-progressive_growing_of_gans/model.py\#L21}
in project GANS (described in \autoref{sec:experimental_subjects});
the first function argument \Py{net}
expects objects encoding Keras network architectures.
This complex type is not directly supported by \an's built-in annotations;
thus, we should define a custom generator function \Py{generator_G_convblock}
that builds valid instances of the type.

To this end, we first look for any \emph{client code} of \Py{G_convblock}.
Another function \Py{Generator} in project GANS, shown in \autoref{lst:generator-example-part2},
calls \Py{G_convblock} (line~\ref{l:gen2:call})
after building a suitable network architecture object
(line ~\ref{l:gen2:net}).
Thus we can use parts of \Py{Generator} to build \Py{generator_G_convblock}:
the ``extract function''
refactoring\projURL[Extract function refactoring]{https://refactoring.com/catalog/extractFunction.html}
applied to lines \ref{l:gen2:first}--\ref{l:gen2:last} in \autoref{lst:generator-example-part2}
outputs \autoref{lst:generator-example-part3}'s generator function.
Now, \Py{generator_G_convblock} is a new function, which we can annotate like any other functions
that is processed by \annotest.

In this example
it was easy to identify a contiguous sequence of statements
and extract it into a generator function.
In other cases, the relevant client code may mix statements useful for the generator
with others that pertain to a different functionality.
In these cases, we can simply extract a larger snippets of code,
and then refactor it to remove unused statements.
In \autoref{lst:generator-example-part2}'s example,
we could extract all lines \ref{l:gen2:start}--\ref{l:gen2:last} into a new function;
then, all statements before line~\ref{l:gen2:first}
are not used by the final line~\ref{l:gen2:last},
and thus can be removed from the generator
(leading to the same generator as in \autoref{lst:generator-example-part3}).
In all the experiments of this paper,
these simple refactoring steps were sufficient to build
all necessary custom generator functions.

\begin{lstlisting}[float=*,caption={Signature of project GANS's function \Py{G_convblock}, whose first argument \Py{net} requires a custom generator.}, label={lst:generator-example-part1},language=aN]
@arg(net): objs(generator_G_convblock)
def G_convblock(net, num_filter, filter_size, actv, init,
                pad='same', use_wscale=True, use_pixelnorm=True, use_batchnorm=False, name=None):
   # ... 24 lines of body code ...
\end{lstlisting}

\begin{lstlisting}[float=*,caption={An excerpt of project GANS's function \Py{Generator}, a client of \autoref{lst:generator-example-part1}'s function \Py{G_convblock}.}, label={lst:generator-example-part2},language=aN]
def Generator(num_channels=1, resolution=32, label_size=0, fmap_base=4096, fmap_decay=1.0,
              fmap_max=256, latent_size=None, normalize_latents=True, use_wscale=True, 
              use_pixelnorm=True, use_leakyrelu=True, use_batchnorm=False, tanh_at_end=None,
              **kwargs):
    R = int(np.log2(resolution))   (*\label{l:gen2:start}*)
    assert resolution == 2 ** R and resolution >= 4   (*\label{l:gen2:assert}*)
    cur_lod = K.variable(np.float32(0.0), dtype='float32', name='cur_lod')

    def numf(stage): return min(int(fmap_base / (2.0 ** (stage * fmap_decay))), fmap_max)
    if latent_size is None:
        latent_size = numf(0)
    (act, act_init) = (lrelu, lrelu_init) if use_leakyrelu else (relu, relu_init)

    inputs = [Input(shape=[latent_size], name='Glatents')](*\label{l:gen2:first}*)
    net = inputs[-1]

    if normalize_latents:
        net = PixelNormLayer(name='Gnorm')(net)
    if label_size:
        inputs += [Input(shape=[label_size], name='Glabels')]
        net = Concatenate(name='G1na')([net, inputs[-1]])
    net = Reshape((1, 1,K.int_shape(net)[1]), name='G1nb')(net)   (*\label{l:gen2:net}*)(*\label{l:gen2:last}*)

    net = G_convblock(net, numf(1), 4, act, act_init, pad='full', use_wscale=use_wscale, (*\label{l:gen2:call}*)
                      use_batchnorm=use_batchnorm, use_pixelnorm=use_pixelnorm, name='G1a')
    # ... 20 more lines of code ...
\end{lstlisting}

\begin{lstlisting}[float=*,caption={The custom generator for argument \Py{net} of \autoref{lst:generator-example-part1}'s function \Py{G_convblock}, built by factoring out lines \ref{l:gen2:first}--\ref{l:gen2:last} in \autoref{lst:generator-example-part2}.}, label={lst:generator-example-part3},language=aN]
@generator
@exclude
@arg(latent_size): ints(min=1)
@arg(normalize_latents): bools()
@arg(label_size): ints()
def generator_G_convblock(label_size, latent_size, normalize_latents):
    inputs = [Input(shape=[latent_size], name='Glatents')]
    net = inputs[-1]
    if normalize_latents:
        net = PixelNormLayer(name='Gnorm')(net)
    if label_size:
        inputs += [Input(shape=[label_size], name='Glabels')]
        net = Concatenate(name='G1na')([net, inputs[-1]])
    net = Reshape((1, 1, K.int_shape(net)[1]), name='G1nb')(net)
    return net
\end{lstlisting}

\subsection{Test Generation}
\label{sec:test_generation}

The annotations written in the \an language
supply all the information that is needed
to generate unit tests for every annotated function.
In principle, we could use
any technique for test-case generation
and then filter any generated tests, keeping only
those that comply with the annotations.
However, the experiments reported in \autoref{sec:rq4-answer} indicate
that such an aimless strategy would be inefficient,
especially given the dynamically typed nature of Python.

Instead, \annotest uses \emph{property-based test-case generation} to actively match the constraints introduced
by \an annotations.
More precisely, the current implementation of \annotest
uses the Hypothesis property-based
test-case generator\projURL[Hypothesis: ]{https://hypothesis.readthedocs.io}
through its API.
To test a Python function using Hypothesis,
we have to write a \emph{test template}, which consists
of a \emph{parametric} unit test method
that calls a collection of \emph{strategies}.
A strategy is a sort of generator function, 
which outputs values of a certain kind.
A parametric test method calls some of the strategies,
combines their outputs, and uses them to call
the function under test.

\begin{lstlisting}[float=*,caption={Hypothesis test template built by \annotest for \Py{DenseNet}'s annotations in \autoref{lst:motivational1_annotation}.}, label={lst:motivational1_hypothesis}]
@given(input_shape=tuples(integers(min_value=20, max_value=70), (*\label{l:hy:integers}*)
                          integers(min_value=20, max_value=70),
                          integers(min_value=1, max_value=3)),
       dense_blocks=integers(min_value=2, max_value=5),
       dense_layers=one_of(st.just(-1),
                         integers(min_value=1, max_value=5),
                         int_lists_an(min_len=2, max_len=5, min=2, max=5)), (*\label{l:hy:int_lists_an}*)
       growth_rate=integers(min_value=1, max_value=20),
       nb_classes=integers(min_value=2, max_value=22),
       dropout_rate=floats(min_value=0, max_value=1, (*\label{l:hy:floats}*)
                           exclude_min=True, exclude_max=True),
       bottleneck=booleans(),
       compression=floats(min_value=0, max_value=1, exclude_min=True),
       weight_decay=floats(min_value=0.0001, max_value=0.01),
       depth=integers(min_value=10, max_value=100))
@settings(deadline=None, suppress_health_check=[HealthCheck.filter_too_much,
                                                HealthCheck.too_slow])
def test_DenseNet(input_shape, dense_blocks, dense_layers, growth_rate,
                  nb_classes, dropout_rate, bottleneck, compression,
                  weight_decay, depth):
    assume(type(dense_layers) is not list or (*\label{l:hy:assume}*)
           len(dense_layers) == dense_blocks)
    DenseNet(input_shape, dense_blocks, dense_layers, growth_rate, (*\label{l:hy:densenet}*)
             nb_classes, dropout_rate, bottleneck, compression,
             weight_decay, depth)
             
@defines_strategy() (*\label{l:hy:int_lists_st_begin}*)
def int_lists_an(min_len=1, max_len=None, min=1, max=None):
    if max_len is None:
        max_len = min_len + 2
    if max is None:
        max = min + 5
    return lists(integers(min, max), 
                 min_size=min_len, max_size=max_len)  (*\label{l:hy:int_lists_st_end}*)
\end{lstlisting}

\annotest automatically builds a suitable Hypothesis strategy
for each \An{@arg} annotation.
Hypothesis provides built-in strategies
that cover basic type annotations,
such as Python's atomic types and tuples.
\annotest reuses the built-in strategies whenever possible,
and combines them to generate values for more complex
or specialized constraints (such as \An{int_lists}).
For instance,
\autoref{lst:motivational1_hypothesis}
shows parts of the parametric tests
generated by \annotest to encode the annotations 
in \autoref{lst:motivational1_annotation}'s running example.
\annotest reuses 
Hypothesis's built-in strategies
\Py{integers} (line~\autoref{l:hy:integers})
and \Py{floats} (line~\autoref{l:hy:floats}); %
and combines
Hypothesis strategies \Py{lists} and \Py{integers}
(lines~\ref{l:hy:int_lists_st_begin}--\autoref{l:hy:int_lists_st_end})
to render \an's \An{int_lists} type constraint.

To encode arbitrary \An{objs} annotations
(\autoref{sec:annotations-generators}),
\annotest first builds strategies for 
the annotations of 
each user-written custom generator function,
as if it was testing the generator;
then, it combines them
to build a new strategy that follows
the generator's implementation to
output the actual generated objects---%
used as inputs for the function under test.

To encode \An{@require} annotations
(preconditions),
\annotest uses Hypothesis's \Py{assume} function.
When test-case generation
reaches an \Py{assume}, it checks whether
its Boolean argument evaluates to true:
if it does, generation continues as usual;
if it does not, the current test input is discarded,
and the process restarts with a new test.
Thus, \Py{assume}s can effectively act as filters
to further discriminate between test inputs---%
a feature that \annotest leverages to
enforce precondition constraints 
where appropriate in a parametric test.
Line~\ref{l:hy:assume} in \autoref{lst:motivational1_hypothesis}
shows an example of using \Py{assume}
to encode the running example's precondition (line~\ref{l:an:require} in \autoref{lst:motivational1_annotation}).

After translating the annotations
into suitable test templates,
\annotest simply runs Hypothesis on those templates.
The property-based test-case generator
``runs'' the templates to
build unit tests that satisfy the encoded properties;
it also runs these unit tests, and reports any failure to
the user.
Hypothesis's output is also \annotest's final output
to the user.

\nicepar{Alternative back-ends.}
\annotest's current implementation uses 
Hypothesis as back-end, 
since property-based testing is a framework for defining testing properties
in a naturally \emph{generative} way.
However,
using other test-input generation engines as back-end
is possible in principle.
Automatically translating all \an annotations to preconditions (Boolean predicates) is straightforward,
which could be passed to a tool like Deal~\cite{Deal}.
As we demonstrate in \autoref{sec:rq4-answer},
Deal is not very effective at \emph{generating} inputs
that satisfy the preconditions, when these encode the complex combinations of constraints
that are common in NN programs;
however, Deal can also use preconditions for \emph{static checking},
which would provide a complementary usage of \annotest's annotations.
Pynguin~\cite{Lukasczyk:2020} is a general-purpose test-case generator for Python.
In order to use it as a back-end for \annotest,
we could leverage its genetic algorithm, which tries to maximize the \emph{branch coverage}
of the tests it generates.
As done with EvoSuite (a test-case generation tool for Java that is also based on genetic algorithms)
in related work~\cite{DBLP:journals/ese/FraserA15,GFMFZ-TSE15-DynaMate},
one could express the input constraints as a series of branches in the instrumented program,
so that Pynguin would be driven to find inputs that ``pass'' all the constraints%
---the valid inputs that we are looking for.

\subsection{Failing Tests and Oracles}
\label{sec:oracle_problem}

The \annotest approach, and the \an annotation language on which it is based,
works independent of how a test is classified as failing or passing.
In other words, \annotest generates test inputs that are consistent with the annotations;
determining whether the resulting program behavior is correct
requires an \emph{oracle}~\cite{Barr:2015}.
In this paper, we only ran the tests generated by \annotest
with \emph{crashing} oracles:
an execution is \emph{failing} when it cannot
terminate normally, that is it leads to an assertion violation,
an unhandled exception, or some other low-level abrupt termination.

While crashing bugs are the most frequent ones,
NN programs also exhibit other kinds of bugs such as
performance loss, data corruption, and incorrect output~\cite{Islam:2019}.
In principle, if we equipped the NN programs with oracles suitable to detect
such kinds of bugs, \annotest could still be used to generate
test inputs.
However, some of these bug categories may be easier to identify by testing
a NN at a different level than the bare program code.
For example, bugs that lead to poor robustness of a NN classifier
involve testing a fitted model rather than the model's implementation~\cite{Sun:2019,Hu:2019,Shen:2018}.
Revisiting the \annotest approach to make it applicable to different kinds of oracles
belongs to future work.

\section{Experimental Evaluation}
\label{sec:experiments}

The experimental evaluation 
aims at determining whether the \annotest
approach is effective at detecting real bugs in NN programs,
and whether it requires a reasonable annotation effort.
Precisely, we address the following research questions:

\begin{description}[leftmargin=9.8mm]
\item[RQ1.] Does \annotest generate tests that expose bugs with few false positives (invalid tests)?
            
\item[RQ2.] Can \annotest reproduce known, relevant bugs (that were
  discovered and confirmed by expert manual analysis)?

\item[RQ3.] How many annotations does \annotest need to be effective?

\item[RQ4.] How does \annotest compare to other generic (non-NN
  specific) test-case generation techniques?
  
\item[RQ5.] How does \annotest compare to manual-written tests in terms of coverage?
\end{description}

\subsection{Experimental Subjects}
\label{sec:experimental_subjects}

To include a broad variety of real-world NN projects,
we selected our experimental subjects
following Islam et al.~\cite{Islam:2019}'s extensive survey
of bugs and their replication package,\projURL[Islam et al.'s NN bugs dataset:]{https://lab-design.github.io/papers/ESEC-FSE-19/}
which collects hundreds of NN program bugs
from Stack Overflow posts and public GitHub projects.
The former are unsuitable to evaluate \annotest,
since they usually consist of short, often incomplete, 
snippets of code that punctuate a natural-language text.
In contrast, 
the GitHub projects provide
useful subjects for our evaluation. %

The survey~\cite{Islam:2019}
lists 557 bugs in 127 GitHub projects using the NN frameworks Keras, TensorFlow, PyTorch, Theano, and Caffe.
With 350 bugs in 42 projects, 
Keras is the most popular project in this list;
we target it for the bulk of our evaluation.
Starting from all 42 Keras projects,
we excluded:
\begin{enumerate*}[label=\emph{(\roman*)}]
\item 3 projects that were no longer publicly available;
\item 7 projects with no bugs classified as ``crashing''
  (see \autoref{sec:oracle_problem});
  \item and 5 projects that still use Python~2.
\end{enumerate*}
While it could be modified to run with Python~2,
we developed \annotest primarily for Python~3,
which is the
only supported major version of the language at
the time of writing.
We excluded another 4 projects whose
repositories were missing some components necessary
to execute them 
(such as data necessary to train or test the NN model, or
to otherwise run the NN program).
Finally, 7 projects did not include
any reproducible crashing bugs
(see \autoref{sec:experimental_setup} for 
how we determined these).
This left 16 projects using Keras, 
which we selected for our evaluation.
To demonstrate that \annotest
is applicable also to other NN frameworks,
we also selected %
2 projects based on TensorFlow and 1 project based on PyTorch;
these are among the largest projects using those frameworks
analyzed by Islam et al.~\cite{Islam:2019}.
The leftmost columns of
\autoref{tab:reproduction}
list all selected 19 projects used in our evaluation,
and their size in lines of code and number of functions.
These projects (and their known bugs) are based on Islam et al.~\cite{Islam:2019}'s detailed survey of real-world NN bugs; this
ensures that our subjects are representative of realistic NN programs
and of the bugs that commonly affect them.

\nicepar{Comparison with Pynguin.}
To answer RQ4, we want to compare \annotest to Pynguin (a general-purpose test-case generator for Python programs) on generating tests for realistic NN programs.
Unfortunately, all the NN projects that we use for \annotest's evaluation
are incompatible with Python~3.8 (mainly because they require TensorFlow 1.x),
whereas Pynguin only runs with Python~3.8 (or later versions).
Therefore, we considered
PyTorch's machine vision 
project Vision:\projURL[Vision (0.11.2): ]{https://github.com/pytorch/vision/tree/v0.11.2}
an actively maintained open-source NN program that is compatible
with Python 3.8 and includes type hints (used by Pynguin).
Pynguin can only generate tests for 40 of Vision's 104 modules;
current limitations\footnote{Including bugs, one of which we reported to Pynguin's maintainers who fixed it.\projURL*{https://github.com/se2p/pynguin/issues/20}}
of its implementation prevent it from running correctly on the other 64 modules.
For our experiments,
we selected module \Py{mnist}\projURL{https://github.com/pytorch/vision/blob/v0.11.2/torchvision/datasets/mnist.py} in package \Py{torchvision.dataset}---%
one of the largest among those that Pynguin can analyze.

\nicepar{Comparison with manual tests.}
Manually writing \an annotations, and then letting \annotest generate tests automatically, 
is an alternative to the usual approach of writing unit tests manually.
Thus, RQ5 compares
manually-written tests to those generated by \annotest
in terms of coverage.
Unfortunately, none of the 19 projects selected by Islam et al.~\cite{Islam:2019}
contains any unit tests.%
\footnote{Project ADV includes a single integration test; the other projects include no tests at all.}
Therefore, we resorted to project Vision again, as it contains an extensive manually-written test suite.
For our experiments, we selected three Vision modules of substantial size that are tested in different ways:
module \Py{backbone_utils} is among the most thoroughly tested
(the project's test suite reaches 96\% branch coverage);
module \Py{image} is fairly well tested
(79\% branch coverage, which is an average coverage figure among the project's modules);
and module \Py{_video_opt} is scarcely tested
(16\% branch coverage, and is only tested indirectly by the unit tests of other client modules).

\subsection{Experimental Setup}
\label{sec:experimental_setup}

This section describes how we setup each project
before applying \annotest;
and the experiments we conducted to answer the 
RQs.

\subsubsection{Project Setup}
As first step, we created an 
Anaconda\projURL[Anaconda:]{https://www.anaconda.com/}
environment for each project %
to configure and run it independent of the others.
Every project has \emph{dependencies}
that involve specific libraries.
Collecting all required dependencies can be tricky:
a project may work only with certain library versions,
older versions of a library may no longer be available,
and newer backward-compatible versions may conflict
with other dependencies.
A handful of projects detail the specific versions of the libraries they need in a \Py{setup.py}, \Py{requirements.txt}, or Jupiter Notebook file---%
or at least in a human-readable \Py{readme}.
In many cases, none of these were available,
so we had to follow a trial-and-error process:
\begin{enumerate*}[label=\emph{(\roman*)}]
\item search the source code for \Py{import L} statements;
\item retrieve the version of library \Py{L} that was up-to-date
around the time of the project's analyzed commit;
\item in case that version is no longer available or conflicts
with other libraries, try a slightly more recent or slightly older
version of \Py{L}.
\end{enumerate*}

NN programs usually need
\emph{datasets} to run.
When a suitable dataset was not available in a project's repository,
we inspected the source code and its comments 
to find references to public datasets
that could be used, fetched them, and added them to the
project's environment.
In a few cases, the project included functions to generate
a sample dataset, which was usually suitable to be able to at least test the project.
For a few projects using very large datasets, 
we shrank them by removing some data points 
so that certain parts of the project's code
ran more efficiently.
Whenever we did this, we ascertained that 
using the modified dataset did not affect 
general program behavior in terms of \emph{reachability}---%
which is what matters for detecting the crashing
bugs that we target in our evaluation.

Properly setting up all NN programs
so that they can be automatically run and tested
was quite time-consuming at times,
since several of the projects' repositories are
incomplete, outdated, and poorly documented.
Our replication package includes all required dependencies,
which can help support future work in this area.

\subsubsection{Experimental Process}
To address RQ1, we
selected the latest versions
of two projects
among the largest and most popular ones
(ADV and GANS in \autoref{tab:full-annotations})
and followed the guidelines described in \autoref{sec:guidelines}
to fully annotate them with \an.
``Fully annotate'' means that we tried
to annotate every function
of the project's source code, 
and to write annotations
that are as accurate as possible:
neither unnecessarily constraining
(skipping some valid inputs)
nor too weak (allowing invalid inputs).

To address RQ2, 
we tried to use \annotest to reproduce
the bugs reported by Islam et al.~\cite{Islam:2019}
for the selected projects.
More precisely, Islam et al.~\cite{Islam:2019}'s 
companion dataset
identifies each bug $b$
by a triple $(\ell, b^-, b^+)$:
line $\ell$ in commit $b^-$
is the faulty statement,
which is fixed by the (later) commit $b^+$.
As we mentioned above, 
Islam et al.~\cite{Islam:2019}'s dataset
was collected by manual analysis,
and thus some of the bugs are
not (no longer) reproducible, 
are duplicate, 
or are otherwise outside \annotest's scope.
For our evaluation, we selected only 
\emph{unique reproducible crashing bugs}:
\begin{enumerate*}[label=\emph{(\roman*)}]
\item ``crashing'' means that the fault triggers
a runtime program failure, which we use as oracle;\footnote{While Islam et al.~\cite{Islam:2019} classify some bugs as ``crashing'', we also included bugs in other categories provided they can eventually generate a crash.}
the crashing location $c$
may be different from the bug location $\ell$;
\item ``reproducible'' means that we could manually
run the program to trigger the failure;
\item ``unique'' means that we merged bugs that are
indistinguishable by a crashing oracle 
(for example, they crash at the same program point, 
or they fail the same assertion) 
or that refer to the very same triple in Islam et al.~\cite{Islam:2019}'s dataset.
\end{enumerate*}
 
Out of all 213
bugs in Islam et al.~\cite{Islam:2019} for the 19 selected projects,
we identified 81 unique reproducible crashing bugs.
For each such bug $b = (\ell, c, b^-, b^+)$
we annotated the project's commit $b^-$
starting from the function (or method) \Py{f} where location $\ell$ is,
and continuing with the other functions that
depend on \Py{f}.
We stopped annotating as soon as
the annotations where sufficient to exercise
function \Py{f} (including, in particular, reaching $\ell$ and/or crash location $c$).
Then, we ran \annotest to generate
tests for \Py{f} and any other functions that we annotated.
We count bug $b$ as \emph{reproduced}
if some of the generated tests
fails at crashing location $c$,
and doesn't fail if run on the patched version~$b^+$.

To address RQ3,
we measured the annotations we wrote for RQ1's and RQ2's experiments;
and we compared the size (in lines of code) of
these manually-written annotations to the Hypothesis
code generated automatically by \annotest from the annotations.

To address RQ4,
we compared
\annotest to Pynguin and Deal.
As we discuss in \autoref{sec:related-work},
Pynguin~\cite{Lukasczyk:2020} is a state-of-the-art unit-test generator for Python
that uses type hints to improve its effectiveness (although it also works without type hints);
Deal~\cite{Deal} is a Python library for Design by Contract, supporting annotations such as preconditions, as well as test-case generation and static analysis based on them.
For the comparison with Pynguin,
we annotated the functions in Vision's module \Py{mnist} (see \autoref{sec:experimental_subjects})
using \an similarly to what done for RQ1, writing 21 regular annotations and 1 generator for 23 functions under test;
then, we compared Pynguin's generated tests to \annotest's.
For the comparison with Deal,
we took all functions
in our running examples \autoref{lst:motivational1}--\ref{lst:example_cc_example} %
and added preconditions in Deal's syntax
that express the same input constraints
as our annotations in \an's syntax;
then, we compared Deal's generated tests to \annotest's.

To address RQ5,
we annotated the functions in Vision's modules
\Py{backbone_utils}, \Py{image}, and \Py{_video_opt}
(see \autoref{sec:experimental_subjects})
using \an similarly to what done for RQ1.
Since the goal is comparing to manually written tests,
we ignored the tests when writing \an annotations,
and only considered examples of function usages in the library implementation or comments.
Using tool \emph{Coverage.py}\projURL[Coverage.py v.~6.5.0]{https://github.com/nedbat/coveragepy}
we measured the branch coverage achieved on each module by:
\begin{enumerate*}[label=\emph{(\roman*)}]
\item the manually-written unit tests in Vision's test suite;
\item the tests generated by \annotest from the annotations.
\end{enumerate*}
We used branch coverage but note that, on these subjects, this metric correlates very strongly
(Pearson correlation coefficient: 0.94) with statement coverage;
thus, using either coverage metric would lead to the same findings.

\nicepar{Annotation effort.}
As we mentioned in \autoref{sec:guidelines}, 
gaining an accurate understanding of a program's behavior
is necessary regardless of the approach one follows to build tests.
In our experiments, we found that
finding plausible ranges for a function's inputs
requires only modest effort in the majority of cases.
This is in accordance with the so-called \emph{locality principle}~\cite{locality-story},
which implies that a significant part of a program's behavior often can be understood by observing
only a small number of program inputs~\cite{locality}.
Regardless of whether one is targeting a program that is easy or hard to test,
\annotest can support the tester's job
by providing a means of expressing the input constraints,
of exercising them with automatic test generation.

\subsection{Experimental Results}
\label{sec:experimental_results}

\begin{table*}[!tb]
  \centering

  \begin{tabular}{l r r rrr rrr}
    \toprule
    \multicolumn{1}{c}{\textsc{project}}
    & \multicolumn{1}{c}{\textsc{loc}}
    & \multicolumn{1}{c}{\textsc{functions}}      
    & \multicolumn{3}{c}{\textsc{annotations}}      
    & \multicolumn{3}{c}{\textsc{bugs}}
    \\
    \cmidrule(lr){4-6} \cmidrule(lr){7-9}
    & & &
      \multicolumn{1}{c}{\textsc{\#a}} & \multicolumn{1}{c}{\textsc{\%f}} &
      \multicolumn{1}{c}{\textsc{\%g}}
    &
      \multicolumn{1}{c}{\textsc{true}} & \multicolumn{1}{c}{\textsc{spurious}} & \multicolumn{1}{c}{\textsc{precision}}
    \\
    \midrule
    ADV
    & 1421
    & 100
    & 1.58
    & 49\%
    & 7\%
    & 33
    & 5
    & 87\%
    \\
    GANS
    & 2496
    & 149
    & 1.15
    & 37\%
    & 6\%
    & 17
    & 1
    & 94\%
    \\
    \cmidrule(lr){2-9}
    \multicolumn{1}{c}{\textbf{overall}}
    & 3917 & 249 &
    1.33 & 42\% & 7\% 
    & 50 
    & 6 
    & 89\%
    \\
    \bottomrule
  \end{tabular}
  \caption{Two projects fully annotated with \annotest and the found bugs.
    Each row shows data about a \textsc{project}
    (identified by an acronym; see \autoref{tab:reproduction} 
     for the URL of their GitHub repositories):
    its size in lines of code \textsc{loc}
    and number of \textsc{functions} (including methods);
    the average (per function) number~\textsc{\#a} of annotations we added to the project, 
    the percentage~\textsc{\%f} of functions with at least one annotation,
    and the percentage~\textsc{\%g} of annotations
    that use custom generators;
    and the number of unique crashing \textsc{bugs} found by generating
    tests based on the templates---split into confirmed \textsc{true} bugs,
    \textsc{spurious} bugs (triggered by invalid inputs),
    and the corresponding $\textsc{precision} = \textsc{true}/(\textsc{true}+\textsc{spurious})$.}
  \label{tab:full-annotations}
\end{table*}

\subsubsection{RQ1: Precision}
\label{sec:rq1-answer}

\autoref{tab:full-annotations} shows the results of applying \annotest
to the latest commits\footnote{The projects are however no longer maintained; therefore, we did not submit any of the found bugs to the projects' repositories.} of projects ADV and GANS.
With the goal of annotating the projects
as thoroughly as possible,
we ended up writing some \an annotations
for 42\% of their 249~functions.
Most of the functions that we left without annotations
do not need any special constraints to be tested---%
usually because they either are simple utility functions
that are only called in specific ways by the rest of the project or have no arguments.
There are a few additional cases of functions
that are not used anywhere in the project and
whose intended usage we could not figure out
in any other way;
in these cases, we did not annotate them 
(and excluded them from testing).
With these annotations, \annotest
reported 
56
crashes, 
50
of which
we confirmed as genuine unique crashing bugs;
this corresponds to a precision of 
89\%.

As previously reported~\cite{Sun:2017}, bugs due to project dependency conflicts are quite common in NN programs.
An interesting example is a crash that
occurs in ADV when it accesses
attribute \Py{W}\projURL{https://github.com/bstriner/keras-adversarial/blob/master/examples/example_rock_paper_scissors.py#L62} in Keras's class 
\Py{Dense}.\projURL{https://github.com/keras-team/keras/blob/keras-1/keras/layers/core.py#L588}
This attribute was renamed to \Py{kernel}\projURL{https://github.com/keras-team/keras/blob/keras-2/keras/layers/core.py#L823}
in Keras version 2.0.
Since ADV explicitly supports this major version of Keras,
this crash is a true positive.
Another confirmed bug we found 
was due to a function in ADV still using 
tuple parameter unpacking\projURL{https://github.com/bstriner/keras-adversarial/blob/master/examples/example_aae_cifar10.py#L69-L70}---%
a Python~2 feature removed in Python~3.
The ADV project developers probably forgot to
update this one instance consistently with how 
they updated the rest of the project,\projURL{https://github.com/bstriner/keras-adversarial/blob/master/examples/example_aae.py#L46-L47} 
which is indeed designed to work with Python~3.

A tricky example of false positive
occurred in project GANS's function
\Py{create_celeba_channel_last},\projURL{https://github.com/naykun/TF_PG_GANS/blob/master/Tensorflow-progressive_growing_of_gans/h5tool3.py\#L500}
which creates an HDF5\projURL[HDF5 (Hierarchical Data Format 5) for Python:]{https://www.h5py.org/} file for the CelebA dataset~\cite{Liu:2015}.
One of the tests generated by \annotest
crashes\projURL{https://github.com/naykun/TF_PG_GANS/blob/master/Tensorflow-progressive_growing_of_gans/h5tool3.py\#L520}
as it is unable to create a file.
However, the failure does not happen
if we run the function manually
using the very same inputs;
thus, the testing environment is responsible for the spurious failure.

These experiments suggest that \annotest
can be quite effective to pin down
bugs, problems, and inconsistencies
in NN programs, thus helping systematically
improve their quality.

\begin{result}
Applied to two fully-annotated open-source NN programs,
\annotest generated tests revealing 
50~bugs with 
89\%~precision.
\end{result}

\subsubsection{RQ2: Recall}
\label{sec:rq2-answer}

\autoref{tab:reproduction} shows the results 
of applying \annotest 
to detect
81 unique reproducible crashing bugs in 19 projects
surveyed by Islam et al.~\cite{Islam:2019}
and selected as explained in \autoref{sec:experimental_subjects}.
Using the annotations we provided,
\annotest reproduced 63 of these bugs
without generating any spurious failing tests.
This corresponds to a 100\% precision
and 78\% recall relative to the
unique reproducible known bugs from Islam et al.~\cite{Islam:2019}.
With the same annotations,
\annotest also revealed another 31
failures that we confirmed as additional
crashing bugs in the same projects.\footnote{Islam et al.~\cite{Islam:2019}'s survey is not meant to be an exhaustive catalog of all bugs in these projects.}

While \annotest was quite effective at reproducing 
the known bugs in these projects,
it's interesting to discuss the issues
that prevented it from achieving 100\% recall.
We identified several scenarios:
\begin{enumerate*}[label=\emph{(\roman*)}]
\item masking;
\item scripting code;
\item nested functions;
\item lazy features;
\item and inaccessible code.
\end{enumerate*}

\begin{table*}[!tb]
  \centering
  \setlength{\tabcolsep}{4.5pt}
  \begin{tabular}{l l r rr r rrr rrrrrr}
    \toprule
    \multicolumn{1}{c}{} %
    & \multicolumn{1}{c}{\textsc{project}}
    & \multicolumn{1}{c}{\textsc{loc}}
    & \multicolumn{2}{c}{\textsc{functions}}      
    & \multicolumn{1}{c}{\textsc{rev}}
    & \multicolumn{3}{c}{\textsc{annotations}}      
    & \multicolumn{6}{c}{\textsc{bugs}}
    \\
    & & &
    \multicolumn{1}{c}{\textsc{total}} &
    \multicolumn{1}{c}{\textsc{tested}} 
    & & 
      \multicolumn{1}{c}{\textsc{\#a}} & \multicolumn{1}{c}{\textsc{\%f}} &
      \multicolumn{1}{c}{\textsc{\%g}}
    &
      \multicolumn{1}{c}{\textsc{known}}
      & \multicolumn{1}{c}{\textsc{rep}}
      & \multicolumn{1}{c}{\textsc{other}}
      & \multicolumn{1}{c}{\textsc{spurious}}
      & \multicolumn{1}{c}{\textsc{precision}}
      & \multicolumn{1}{c}{\textsc{recall}}
    \\
    \midrule
    K
    & NAAS\projURL{https://github.com/anastassia-b/neural-algorithm-artistic-style}
    & 140	& 7	& 0 & 2	& -- & 0\%	& 0\%	& 2	& 2	& 1	& 0	& 100\% & 100\%
    \\
    K & ADV\projURL{https://github.com/bstriner/keras-adversarial}
    & 1421	& 100	& 4 & 2	& 1.5	& 4\%	& 0\%	& 8	& 6	& 3	& 0	& 100\% & 75\%
    \\
    K & DN\projURL{https://github.com/cmasch/densenet}
    & 82	& 5	& 2 & 1	& 14.0	& 40\%	& 0\%	& 2	& 2	& 2	& 0	& 100\%	& 100\%
    \\
    K & DCF\projURL{https://github.com/csvance/deep-connect-four}
    & 748	& 35	& 1 & 1	& 4.0	& 3\%	& 0\%	& 1	& 0	& 0	& 0	& --	& 0\%
    \\
    K & KIS\projURL{https://github.com/dhkim0225/keras-image-segmentation}
    & 2050	& 92	& 2 & 1	& 1.5	& 2\%	& 0\%	& 6	& 5	& 0	& 0	& 100\% & 83\%
    \\
    K & FRCNN\projURL{https://github.com/dishen12/keras_frcnn}
    & 1643	& 55	& 3 & 1	& 1.7	& 5\%	& 0\%	& 6	& 3	& 0	& 0	& 100\%	& 50\%
    \\
    K & CONV\projURL{https://github.com/heuritech/convnets-keras}
    & 350	& 20	& 0 & 1	& --	& 0\%	& --	& 1	& 0	& 0	& 0	& -- & 0\%
    \\
    K & mCRNN\projURL{https://github.com/jamesmf/mnistCRNN}
    & 225	& 1	& 0 & 1	& --	& 0\%	& 0\%	& 1	& 1	& 5	& 0	& 100\%	& 100\%
    \\
    K & IR\projURL{https://github.com/javiermzll/Image-Recognition}
    & 306	& 38	& 0 & 1	& --	& 0\%	& --	& 2	& 0	& 0	& 0	& -- & 0\%
    \\
    K & RE\projURL{https://github.com/katyprogrammer/regularization-experiment}
    & 966	& 25	& 1 & 1	& 15.0	& 4\%	& 0\%	& 1	& 1	& 5	& 0	& 100\%	& 100\%
    \\
    K & CAR\projURL{https://github.com/michalgdak/car-recognition}
    & 353	& 21	& 1 & 1	& 7.0	& 5\%	& 0\%	& 1	& 1	& 1	& 0	& 100\% & 100\%
    \\
    K & GANS\projURL{https://github.com/naykun/TF_PG_GANS}
    & 2496	& 149	& 2 & 1	& 12.5	& 1\%	& 4\%	& 6	& 4	& 5	& 0	& 100\%	& 67\%
    \\
    K & KAX\projURL{https://github.com/notem/keras-alexnet}
    & 227	& 15	& 0 & 1	& --	& 0\%	& --	& 1	& 0	& 0	& 0	& --	& 0\%
    \\
    K & VSA\projURL{https://github.com/Spider101/Visual-Semantic-Alignments}
    & 630	& 38	& 2 & 1	& 6.0	& 5\%	& 0\%	& 2	& 2	& 4	& 0	& 100\%	& 100\%
    \\
    K & UN\projURL{https://github.com/taashi-s/UNet_Keras}
    & 440	& 28	& 3 & 2	& 3.3	& 11\%	& 30\%	& 6	& 2	& 1	& 0	& 100\% & 33\%
    \\
    K & LSTM\projURL{https://github.com/yagotome/lstm-ner}
    & 477	& 27	& 0 & 1	& --	& 0\%	& --	& 1	& 0	& 0	& 0	& -- & 0\%
    \\
    F & TC\projURL{https://github.com/dennybritz/cnn-text-classification-tf}
    & 285	& 7	&  0 & 2	& --	& 0\%	& 0\%	& 9	& 9	& 2	& 0	& 100\% & 100\%
    \\
    F & TPS\projURL{https://github.com/iwyoo/tf_ThinPlateSpline}
    & 286	& 2	& 2 & 1	& 4.0	& 100\%	& 87\%	& 24	& 24	& 0	& 0	& 100\% & 100\%
    \\
    T & DAF\projURL{https://github.com/zzsdsgdtc/BiDAF_PyTorch}
    & 1094	& 70	& 1 & 1	& 9.0	& 1\%	& 67\%	& 1	& 1	& 2	& 0	& 100\% & 100\%
    \\
    \cmidrule(lr){2-15}
    \multicolumn{2}{c}{\textbf{overall}}
    & 14219	& 735	& 24 & 23	& 6.0	& 3\%	& 12\%	& 81	& 63	& 31	& 0	& 100\%	& 78\%
    \\
    \bottomrule
  \end{tabular}
  \caption{Bugs from Islam et al.~\cite{Islam:2019} that \annotest could reproduce.
    Each row shows data about a \textsc{project} 
    (identified by an acronym and the URL of its GitHub repository):
    its DNN framework (\underline{K}eras, Tensor\underline{F}low, \underline{T}orch),
    its size in lines of code \textsc{loc}
    and the number of \textsc{total} and \textsc{tested} functions (including methods);
    the number of its different \textsc{rev}isions that we analyzed,
    the average (per tested function) number~\textsc{\#a} of annotations we added, 
    the percentage~\textsc{\%f} of functions with at least one annotation,
    and the percentage~\textsc{\%g} of annotations
    that use custom generators;
    and the number of crashing \textsc{bugs} found by generating
    tests based on the templates---the number of \emph{reproducible}
    \textsc{known} bugs reported by Islam et al.~\cite{Islam:2019},
    how many of these the tests \textsc{rep}roduced,
    how many \textsc{other} confirmed true bugs
    and \textsc{spurious} bugs (triggered by invalid inputs)
    the tests also reported in the same experiments,
    and the corresponding $\textsc{precision} = (
    \textsc{rep}
    + \textsc{others})/(
    \textsc{rep}
    + \textsc{others}+\textsc{spurious})$ and $\textsc{recall} = 
    \textsc{rep}
    /\textsc{known}$.
    }
  \label{tab:reproduction}
\end{table*}

\emph{Masking}
occurs when an earlier crash
prevents program execution
from reaching the 
location
of another bug $b'$.
Masking is usually not a problem when
the earlier crash is determined by a known bug $b$:
in this case, we can just run tests 
on the project commit $b^+$ where $b$ has been fixed,
so that execution can reach the other bug $b'$.
However, if a bug $b'$
is masked by an unknown bug
(column \textsc{other} in \autoref{tab:reproduction}),
and we don't know how to fix the unknown bug to allow the program
to continue, $b'$ is effectively unreachable.
We could not reproduce 4 known bugs because of masking.
One of them 
occurrs\projURL{https://github.com/naykun/TF_PG_GANS/commit/efc6c3681587319c72e0e867b2b0e673aa018c17\#diff-2add825310f36eb8852870389321d3e6a7416fed8f9aacd3e0b29fd0a2336b1dL196-L197}
in project GANS,
and is masked by
an unexpected crash\projURL{https://github.com/naykun/TF_PG_GANS/commit/efc6c3681587319c72e0e867b2b0e673aa018c17\#diff-2add825310f36eb8852870389321d3e6a7416fed8f9aacd3e0b29fd0a2336b1dL187}
occurring 
in the same function \Py{Discriminator}.
In project UN, 
some missing statements
make it impossible to distinguish three known bugs,\projURL{https://github.com/taashi-s/UNet_Keras/commit/fd81da67bfcf173331e03687425040138e76bc8f\#diff-e1afe2b6eb4252b0f813153018d4e40a721ed0bac509ce0a3f75d14c046fc800R51}$^,$\projURL{https://github.com/taashi-s/UNet_Keras/commit/fd81da67bfcf173331e03687425040138e76bc8f\#diff-e1afe2b6eb4252b0f813153018d4e40a721ed0bac509ce0a3f75d14c046fc800R52}$^,$\projURL{https://github.com/taashi-s/UNet_Keras/commit/fd81da67bfcf173331e03687425040138e76bc8f\#diff-e1afe2b6eb4252b0f813153018d4e40a721ed0bac509ce0a3f75d14c046fc800R53}
since they all crash the same test.
Therefore, we consider 1 of them reproduced and 2 not reproduced due to masking.
One of the tests produced for project GANS stopped before finding a known bug,\projURL{https://github.com/naykun/TF_PG_GANS/commit/efc6c3681587319c72e0e867b2b0e673aa018c17\#diff-2add825310f36eb8852870389321d3e6a7416fed8f9aacd3e0b29fd0a2336b1dL35}
with a \texttt{SIGKILL} (triggered by memory-related issues).

\annotest
generates unit tests,
which target specific functions in a program's source code.
This excludes any code snippets
in the ``main'' section of a Python file
(under \Py{if __name__=='__main__'}),
which executes when the file is run as a script from 
the command line.
Therefore, \annotest could not reproduce 6
bugs affecting this \emph{scripting code},
such as one known bug in project CONV.\projURL{https://github.com/heuritech/convnets-keras/commit/b1b472ccf59bfc3edb7ad033299875c905bf8e37\#diff-4a9f068fbd6ab76d347ca7772f3da3f100db338cd6c8fb3900adef38ab9dff20L325}
Another example is the only known bug\projURL{https://github.com/notem/keras-alexnet/commit/94638c596ca6f3f474241e8a058fd893e1f5ffaa\#diff-23de837fc8b40e270ddb47d0ae913f55e8d31635b80daa5618273535b9d3cd28L198} in project KAX, which
occurs in a function that depends on command line arguments.

\annotest can test
\emph{nested functions} 
only indirectly, that is when they are
called by a top-level function as part of testing the latter.
It does not support annotating nested functions and generating
unit tests for them since they are not accessible outside their parent functions. 
We could not reproduce 3 known bugs because they affected nested functions.
An example %
is in project FRCNN's function \Py{rpn_loss_regr_fixed_num},\projURL{https://github.com/dishen12/keras_frcnn/commit/d91c0adc5ccd34f6e346fdeddc0a2ce7085a4ffb\#diff-a3429d56d560ec95c6b119754a121d183b32f8a4b73786f8760d083353914efbL18}
which is defined inside top-level function
\Py{rpn_loss_regr}.

  Functions using Python's \Py{yield} statement
  are \emph{lazy}, that is their evaluation is delayed.
  This means that they may not be executed by \annotest's
  unit testing environment (or rather its Hypothesis back-end's).
  We could not reproduce 1 known bug\projURL{https://github.com/dhkim0225/keras-image-segmentation/commit/992685cde39c3d53ea881d22b9cb26e84963d4bb\#diff-d0ff8417443a18c35cc6c3183197d82f48cee72d735133ff901da033d0e32242L89} in project KIS
  because it uses \Py{yield} to build a lazy iterator.
  
  As we remarked above, a bug's crashing location $c$ may differ
  from the actual error location $\ell$ in commit $b^-$.
  If $c$ is in a portion of the code that is \emph{not accessible}
  to the testing environment, \annotest cannot reproduce the bug
  even if it is reproducible in principle.
  This scenario occurred for 3 known bugs that \annotest didn't reproduce.
  Two of them are in project UN\projURL{https://github.com/taashi-s/UNet_Keras/commit/b1b6d938bdd7a3e30f3d1fa58009f4850cbc2958\#diff-e1afe2b6eb4252b0f813153018d4e40a721ed0bac509ce0a3f75d14c046fc800L31}$^,$\projURL{https://github.com/taashi-s/UNet_Keras/commit/b1b6d938bdd7a3e30f3d1fa58009f4850cbc2958\#diff-e1afe2b6eb4252b0f813153018d4e40a721ed0bac509ce0a3f75d14c046fc800L35}
  and only crash in a module whose implementation is incomplete in that program revision.
  Another one\projURL{https://github.com/javiermzll/CCN-Whale-Recognition/commit/e2d3ff925460060f0127c894368147b54b5f03c0\#diff-1b740140b6c82aacc5a6f6b319be9cf103ee72b424ad475f795ea72d4b267849L46} occurs in project IR:
  we tried to no avail to reproduce it at a different, accessible location.
  
  Finally, we could not reproduce 1 bug\projURL{https://github.com/javiermzll/CCN-Whale-Recognition/commit/e2d3ff925460060f0127c894368147b54b5f03c0\#diff-1b740140b6c82aacc5a6f6b319be9cf103ee72b424ad475f795ea72d4b267849L46} in project IR
  simply because we could not figure out
  suitable type constraints to properly exercise the corresponding function.

\begin{result}
\annotest generated tests 
revealing 63 known NN bugs \\
in 19 NN programs,
with a recall of 78\%.
\end{result}

\subsubsection{RQ3: Amount of Annotations}
\label{sec:rq3-answer}

For the \annotest approach to be practical,
it is important that it requires a reasonable
amount of manual annotations.
We leave to future work
a detailed empirical evaluation of the time
and expertise that is needed to write \an annotations.
Here, we discuss quantitative measures of \annotest's annotation
overhead.
We focus on RQ1's experiments (\autoref{sec:rq1-answer}), which analyzed 
projects ADV and GANS in full, as they give a better idea of
the effort needed to use \annotest systematically on whole projects.\footnote{The figures for RQ2's experiments are, however, generally similar.}

\begin{figure}[!bt]
    \centering
    \includegraphics[width=0.47\textwidth]{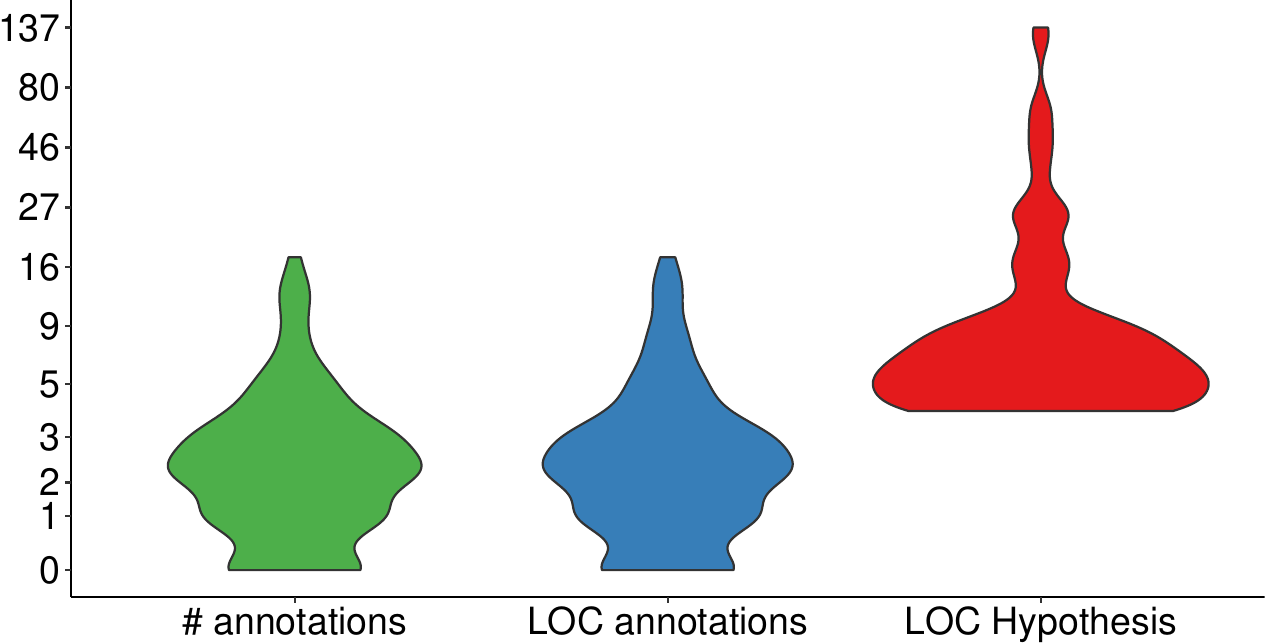}
    \vspace{-3mm}
    \caption{Distribution of the number of \an annotations, lines of code (LOC) of \an annotations,
      and LOC of generated Hypothesis templates for RQ1's experiments.}
    \label{fig:annotations-overhead}
    \vspace{2mm}
\end{figure}

\nicepar{Annotation amount.}
The amount of annotations that we wrote
was usually limited.
In RQ1's experiments, %
we wrote $2$ annotations%
\footnote{An annotation is any instance of the kinds presented in \autoref{sec:an_language}.}
per project function on average (median);
80\% of functions have $3$ annotations or less.
Annotations are mostly concise:
96\% of them fit a single line,
and only 10\% (12) of all functions %
have annotations that span more than 5 lines
(usually decorating functions with several complex arguments).
\autoref{fig:annotations-overhead} pictures the distributions, overall functions, of
number of annotations (left) and lines of code (LOC, middle) of annotations;
since most annotations are a single line, these two distributions are nearly identical.

The  average number of annotations
per tested function %
is higher (6.0) 
in \autoref{tab:reproduction}
since in each of those experiment
we annotated a limited portion of a project
focusing on a specific function
that had a known bug;
therefore, several of the annotations
are duplicated or only slightly modified
from one experiment to the other.
If we had fully annotated the projects,
we would have likely amortized some of this annotation
effort.

In terms of time, we spent, on average, 10--15 minutes to write the annotations of
each function. This time includes inspecting the project's source code to become
familiar with how it works.
As pointed out in \autoref{sec:experimental_setup},
this effort is amortized over various related functions,
and is unevenly distributed, with a few ``complex'' functions
taking considerably more time to understand than most ``simple'' functions.
As mentioned in \autoref{sec:guidelines}, we consider the overall effort comparable to the time
to manually write unit tests for the same functions.

Another way of quantifying the effort-benefit
trade-off is measuring the amount of annotations
per detected bugs: this
ratio is 
6.6 $=330/50$
for the fully-annotated projects in \autoref{tab:full-annotations}
and 
1.5 $=145/94$
for the experiments in \autoref{tab:reproduction}.
These are encouraging figures, if we think
of the amount of manually-written
tests that may have been necessary to discover the same bugs
(see also \autoref{sec:rq5-answer}).

The percentage of annotations using generators 
is higher (12\%) for the projects in \autoref{tab:reproduction}.
More precisely, the two projects in \autoref{tab:full-annotations}
use 15 generators, 73\% (11/15) of which generate NN models.
Among the 16 generators built for the projects in \autoref{tab:reproduction}, 31\% (5/16) generate NN models, 37\% (6/16) provide TensorFlow's tensor objects, and 25\% (4/16) load datasets from disk. The one remaining generator function loads an image from hard disk, turns it into a NumPy array and passes it to a function.
As we explained in \autoref{sec:generator-functions},
we built all generators by applying light refactoring operations to
suitable portions of existing client code within the same project.

\nicepar{Hypothesis overhead.}
Since \annotest translates \an annotations to Hypothesis
templates, we can quantify how concise \an is compared to directly
encoding constraints in Hypothesis.
The rightmost plot in \autoref{fig:annotations-overhead} pictures 
the distribution of LOC of generated Hypothesis code. %
Clearly, Hypothesis code is considerably more verbose than
\an annotations:
Hypothesis templates are 5.5 (median overhead) times longer---11.6 times longer in terms of mean overhead---than the \an annotations they encode,
which points to the benefits of using \an's concise language.

\begin{result}
In our experiments,
\annotest used 2 annotations per function 
on average;
96\% of all annotations fit a single line.
\end{result}

\subsubsection{RQ4: Comparison to Generic Test-Case Generators}
\label{sec:rq4-answer}
We designed \annotest %
not as a general-purpose testing tool but as one specifically geared towards NN programs.
Therefore, we expect \annotest to outperform
generic test-case generators for Python
when generating tests for these programs.

As we discussed in \autoref{sec:experimental_subjects},
we ran \nicepar{Pynguin}
on module \Py{mnist} in project Vision;
the module includes type hints annotations, which Pynguin uses to
improve the accuracy of its generated tests.
Pynguin\footnote{We report experiments that used Pynguin's default configuration; however, using other generation strategies did not significantly change the outcome.} generated 19 tests, %
reporting 6 tests as passing (they terminate without errors),
and 13 tests as failing (they throw an exception).
By manual inspection, we determined that:
\begin{enumerate*}[label=\emph{(\roman*)}]
\item 2 of the 6 passing tests and 10 of the 13 failing tests are
  actually \emph{invalid}, since they call functions with input values
  that are not valid according to the functions' docstring, type hints,\footnote{Pynguin may violate type hints whose format it does not support.}
  or other available documentation;
\item the other 3 failing tests should be classified as \emph{passing}, since
  throwing an exception is the functions' expected behavior in those cases.
\end{enumerate*}
In all, 63\% ($(2+10)/19$) of the tests generated by Pynguin are invalid,
and 79\% ($(3+2+10)/19$) are misclassified.  %
We cannot expect Pynguin to perform better, since it simply lacks the information
to precisely characterize valid inputs;
in contrast, leveraging the \an annotations' information,
\annotest generated 11 tests for module \Py{mnist}: all of them are valid and passing.\footnote{
  While experimenting with testing the Vision project using \annotest,
  we found a bug in a module that Pynguin cannot test.
  We submitted a fix as a pull request\projURL*{https://github.com/pytorch/vision/pull/5238}
  that was promptly accepted.
  Interestingly,
  the affected function\projURL*{https://github.com/pytorch/vision/blob/v0.11.2/torchvision/models/detection/backbone_utils.py#L49} already included a developer-written parameterized test,\projURL*{https://github.com/pytorch/vision/blob/v0.11.2/test/test_backbone_utils.py#L25}
  which nonetheless did not detect ``our'' bug;
  this further demonstrates \annotest's practical effectiveness.
}

\nicepar{Deal}'s
expressive annotation language
is capable of concisely encoding
most of the \an annotations as preconditions (\Py{@deal.pre}).
Then, Deal's test-case generation engine
draws inputs randomly and uses preconditions to filter them;
therefore, the stronger a precondition is,
the more it will struggle to find any valid inputs.
In all our examples (\autoref{lst:motivational1}--\ref{lst:example_cc_example}),
Deal could not generate a single valid input that satisfies all constraints.
Even after removing some of the most complex constraints (for example, the first one in \autoref{lst:motivational1_annotation}), Deal's built-in test-case generator
couldn't generate valid inputs.
Here too, we cannot expect Deal to perform better,
since, unlike \annotest,
its test-case generation process is not built around
the kinds of complex constraints that arise in NN programs.%
\footnote{In addition, Deal focuses on using annotations for static analysis.}

\begin{result}
  \annotest outperforms other test-case generation techniques
  \\
  that are not designed specifically for NN programs.
\end{result}

These results are another manifestation of the 
trade-off between specification accuracy and test effectiveness:
precise tests require precise knowledge of the expected program constraints (and behavior),
regardless of whether this knowledge is formalized as annotations, as executable code,
or is applied directly by the programmer.

\subsubsection{RQ5: Code Coverage}
\label{sec:rq5-answer}

\begin{table*}[!tb]
  \centering
  \begin{tabular}{l rrrrr | rrrr}
    \toprule
    &
    \multicolumn{5}{c}{\textsc{project test suite}}
    & \multicolumn{4}{c}{\textsc{\annotest}}
    \\
    \cmidrule(lr){2-6} \cmidrule(lr){7-10}

    & \multicolumn{2}{c}{\textsc{coverage}}
    & \multicolumn{1}{c}{\textsc{\#functions}} & \multicolumn{2}{c}{\textsc{test size}}
    & \multicolumn{1}{c}{\textsc{\#tests}}
    & \multicolumn{1}{c}{\textsc{coverage}} & \multicolumn{2}{c}{\textsc{annotation size}}
    \\
    \multicolumn{1}{c}{\textsc{module}}
    & \multicolumn{1}{c}{\textsc{unit}}
    & \multicolumn{1}{c}{\textsc{indirect}} 
    & \multicolumn{1}{c}{}
    & \multicolumn{1}{c}{\textsc{loc}}
    & \multicolumn{1}{r}{\textsc{chars}}
    & \multicolumn{1}{c}{\textsc{annotated}}
    & 
    & \multicolumn{1}{c}{\textsc{loc}}
    & \multicolumn{1}{r}{\textsc{chars}}
    \\
    \midrule
    \Py{_video_opt}
    & 0\%
    & 16\%
    & 0
    & 0
    & 0
    & 8
    & 76\%
    & 25
    & 1566
    \\
    \Py{image}
    & 79\%
    & 0\%
    & 15
    & 359
    & 13381
    & 11
    & 84\%
    & 40
    & 1858
    \\
    \Py{backbone_utils}
    & 82\% 
    & 14\% 
    & 9
    & 238
    & 8478
    & 3
    & 82\%
    & 23
    & 1044
    \\
    \midrule
    \multicolumn{1}{c}{\textbf{overall}}
    & 50\% 
    & 5\% 
    & 24
    & 597
    & 21859
    & 22
    & 80\%
    & 88
    & 4468
    \\
    \bottomrule
  \end{tabular}
  \caption{A comparison (part of) Vision's programmer-written test suite and \annotest's generated tests in terms of coverage. For each \textsc{module}, the table reports the branch \textsc{coverage} of the programmer-written \textsc{project test suite} on the module (split between \textsc{unit} tests directly targeting the module's functions, and coverage achieved \textsc{indirect}ly by other modules' tests calling the module); the number of unit test \textsc{functions} directly exercising the module; and the size of these tests in lines of code \textsc{loc} and number of characters \textsc{chars}; the number of functions we \textsc{annotated}; the \textsc{coverage} achieved by \annotest on these functions; and the size of these annotations in lines of code \textsc{loc} and number of characters \textsc{chars}.}
  \label{tab:annotest-vs-test-suite}
\end{table*}

\autoref{tab:annotest-vs-test-suite} compares
the manually-written tests in three of project Vision's modules (see \autoref{sec:experimental_subjects})
to those generated by \annotest after annotating the functions in these modules.

Module \Py{_video_opt} is scarcely tested in Vision:
there are no unit tests for this module (column \textsc{unit} in \autoref{tab:annotest-vs-test-suite}),
but tests in other modules still indirectly exercise 16\% of its branches
(column \textsc{indirect}).
In contrast, \annotest reaches a 76\% coverage after annotating 8 functions in this module.
Vision's unit tests for module \Py{image} achieve a 79\% coverage;
\annotest reaches a higher 84\% coverage.
Finally, Vision's unit tests for module \Py{image} achieve a 82\% coverage, the same
as \annotest.
The whole test suite in Vision actually further exercises module \Py{backbone_utils},
as tests in other modules indirectly add an additional 14\% of coverage.
Overall, \annotest-generated tests achieve a high coverage---comparable to or often higher than that of
the programmer-written test suite.%
\footnote{
  While testing module \Py{image} in these experiments,
  \annotest detected a failure in function \Py{decode_jpeg}\projURL{https://github.com/pytorch/vision/blob/main/torchvision/io/image.py\#L127}
  (which is already thoroughly tested in the project's test suite).
  Reporting this failure\projURL{https://github.com/pytorch/vision/issues/6607}
  to the project maintainers prompted them to modify the function's documentation
  so as to more accurately reflect its intended, implemented behavior.
}

In order to achieve this coverage, what is the amount of code (manual tests) or annotations (\annotest)
that is required?
Vision's unit tests for \autoref{tab:annotest-vs-test-suite}'s three modules
consist of 24 tests, spanning 597 lines of code or 21859 characters;
\annotest's annotations
are only needed for 22 functions,
and span 88 lines or 4468 characters.
This confirms that \annotest's annotations are concise---considerably more concise than unit tests
achieving a lower coverage.
Naturally, the sheer size of a piece of code is an imperfect measure of the effort needed to produce it;
however, \an annotations encode essentially the same information as parametric tests,
and their succinctness is an advantage.

\nicepar{100\% coverage?}
Neither \annotest nor the programmer-written test suite managed to cover 100\% of the branches in the
three modules.
In a few cases, increasing the coverage would be possible by simply writing more unit tests or more general annotations.
For instance, none of the manual tests for module \Py{backbone_utils} instantiates
class \Py{BackboneWithFPN} by passing argument \Py{None}
to its constructor's parameter \Py{extra_blocks};
the corresponding branch\projURL{https://github.com/pytorch/vision/blob/b4686f2b7409d1783dfbb951492cd59bfed08bce/torchvision/models/detection/backbone_utils.py\#L44}
is thus never covered (but it is by \annotest).
Conversely, \annotest does not test
function \Py{_read_video_from_memory}\projURL{https://github.com/pytorch/vision/blob/b4686f2b7409d1783dfbb951492cd59bfed08bce/torchvision/io/_video_opt.py\#L265}
in module \Py{_video_opt} because we could not find meaningful examples of its usage.\footnote{As discussed in \autoref{sec:experimental_setup}, we did not consider the manually-written tests when writing \an annotations, so that the comparison in terms of coverage is fair and meaningful.}
In other cases, however, achieving a 100\% coverage is impractical due to constraints in the test execution environment.
For instance,
a branch\projURL{https://github.com/pytorch/vision/blob/b4686f2b7409d1783dfbb951492cd59bfed08bce/torchvision/io/image.py\#L160}
in function \Py{decode_jpeg} of module \Py{image}
requires running the module on a machine with a GPU supporting the CUDA API.\projURL[CUDA]{https://developer.nvidia.com/cuda-zone}
There is actually a manual test\projURL{https://github.com/pytorch/vision/blob/b4686f2b7409d1783dfbb951492cd59bfed08bce/test/test_image.py\#L382}
covering this branch, but it was not activated in our experiments since we did not run them with CUDA.
We found a few other examples of this scenario\projURL{https://github.com/pytorch/vision/blob/b4686f2b7409d1783dfbb951492cd59bfed08bce/torchvision/io/image.py\#L12}$^,$\projURL{https://github.com/pytorch/vision/blob/b4686f2b7409d1783dfbb951492cd59bfed08bce/torchvision/io/_video_opt.py\#L14}
where increasing the test coverage requires specific hardware or system libraries.

\nicepar{Bug density.}
Users of \annotest write annotations to then generate unit tests automatically. %
In RQ1's experiments, \annotest generated 5649 (valid) tests overall;
only 1\% of them fail and expose a bug.
Thus, bugs in NN programs are \emph{rare}~\cite{rare-bugs}.
This suggests that directly writing tests that selectively expose these bugs may be challenging
even for programmers knowledgeable of the program under test.
The same knowledge is sufficient to write \an annotations and generate tests from them.

\begin{result}
  \annotest achieves high code coverage, \\ comparable to that 
  of manually-written test cases.
\end{result}

\subsection{Threats to Validity}

Identifying
valid test inputs,
and distinguishing between spurious and authentic bugs,
is crucial to ensure \emph{construct validity}
(i.e., the experimental measures are adequate).
Unfortunately, a reliable and complete ground truth
is not available:
the documentation of NN programs is often
incomplete (when it exists), so we had to manually discover
the intended behavior of NN programs from examples,
manual code analysis, and background knowledge.
Our reference---%
Islam et al.~\cite{Islam:2019}'s survey---%
was also compiled by purely manual analysis;
therefore, it does not aim at completeness,
and includes bugs that are not
reproducible (see \autoref{sec:experimental_setup}).
These limitations imply that we cannot make
claims of completeness (``we found all bugs'');
nevertheless, we still have a good confidence
in the correctness of our results (``we found real bugs''):
since we focused on bugs detected 
by crashing oracles, %
most bugs we found with \annotest
are clear violations of
the program's requirements.

Since \annotest uses manually-written annotations,
quantifying the annotation effort
is needed for \emph{internal validity}
(i.e., the experimental results are
suitable to support the findings).
We mostly reported simple measures
(number of annotations, number of functions that require annotations, etc.)
which are unambiguous.
In contrast, we do not make any strong claims
about the time and relative effort needed by
programmers to annotate:
these heavily depend on a programmer's knowledge
of the NN program and of the domain;
precisely assessing them would require controlled experiments
and user studies, which are outside this paper's scope.
However, we remark that expressing \an annotations requires
a knowledge of the program under test of the same kind
that is needed to write effective unit tests.

Picking experimental subjects from
Islam et al.~\cite{Islam:2019}'s extensive
survey of real-world NN bugs helps \emph{external validity}
(i.e., the findings generalize).
As we discussed in \autoref{sec:experimental_subjects},
we excluded some projects for practical reasons
(e.g., no longer available or incomplete)
and we focused on those using the Keras NN framework.
While this focus does not seem especially restrictive
(the majority of projects in the survey uses 
Keras, and we also analyzed projects using other frameworks),
applying \annotest to very different 
kinds of NN programs may require different kinds of
annotations or other changes in the approach.
The \an annotation language is
extensible with
generators (\autoref{sec:annotations-generators}),
which can further help generalizability.
Furthermore, in addition to Islam et al.~\cite{Islam:2019}'s subjects,
we also extensively analyzed the \emph{latest} versions of projects ADV and GANS (\autoref{sec:rq1-answer}),
so that our evaluation did not only include projects with known bugs.

\section{Related work}
\label{sec:related-work}

\nicepar{Automated test-case generation.}
Since testing
is a fundamental activity to ensure software quality~\cite{Candea:2019},
software engineering research has
devised several different techniques to
\emph{automate}
the generation of test inputs~\cite{Anand:2013}. 
Randoop~\cite{Pacheco:2007} (based on random testing)
and EvoSuite~\cite{Fraser:2011} 
(based on genetic algorithms~\cite{Ali:2010})
are 
two of the most popular tools for Java implementing 
automated test-case generation.
Techniques such as those implemented by Randoop and Evosuite
usually depend on the typing information about a method's input
that is provided statically in languages such as Java.

\nicepar{Test-case generation for Python.}
In contrast, programs written in 
dynamically typed languages like Python
do not include such information,
which complicates test-case generation.
In fact, despite Python's popularity~\cite{Cass:2021},
the first widely available tools
for automated test-case generation in Python
appeared only in recent years~\cite{Hypothesis,Lukasczyk:2020}.
Pynguin~\cite{Lukasczyk:2020} 
is based on genetic algorithms like EvoSuite,
and relies on Python's type hints.
Hypothesis~\cite{Hypothesis}
implements property-based testing,
which generates random inputs trying to satisfy
some programmer-written properties.
Deal~\cite{Deal}
is a Python library for design by contact that provides decorations to express pre-
and postconditions; based on them,
it supports both static and dynamic (i.e., test-case generation)
analysis.
\annotest is also an automated test-case generator for
Python, but it provides a specialized set of expressive annotations
useful to precisely express the valid inputs
of NN programs.
Then, it defers the actual test-input generation 
to Hypothesis, which it uses as back-end.
As we demonstrated in \autoref{sec:rq4-answer},
directly using Pynguin, Hypothesis, or Deal to generate
tests for NN programs might be possible in principle,
but it would involve plenty of additional manual work
to express the necessary constraints indirectly through 
a combination of type hints (Pynguin)
and
testing strategies (Hypothesis),
and to program test-case generation strategies that match them (Deal).

\nicepar{Bugs in NN programs.}
Following the increasing in popularity of NN and
other forms of machine learning (ML),
some recent research has looked into the
nature of bugs that occur in NN and ML programs
to understand how they differ compared to
``traditional'' software.
Thung et al.~\cite{Thung:2012} 
studied bugs and human-written patches 
in 3 ML projects (Apache Mahout, Lucene, and OpenNLP)
and classified them according to criteria such as
bug severity and fixing effort. 
A similar study~\cite{Sun:2017} of three other ML projects (Scikit-learn, Paddle, and Caffe) 
revealed that compatibility bugs due to conflicts between project dependencies are quite common in these programs---%
as they were in the subjects we used in \autoref{sec:experiments}'s
experiments.

Zhang et al.~\cite{Zhang:2018}'s analysis
of TensorFlow-based NN projects 
found that 
modeling mistakes, 
incorrect shape of input tensors, 
and unfamiliarity of users with TensorFlow's computation model
were among the most frequent origins of bugs. %
Once again, these findings 
set NN programs apart from traditional software.
Recent studies by Islam et al.~\cite{Islam:2019,Islam:2020} 
on 5 NN frameworks 
confirmed some of Zhang et al.~\cite{Zhang:2018}'s findings
and
further found that bug fix patterns in NN programs are often
different compared to traditional programs. 
In the same line of research,
Humbatova et al.~\cite{Humbatova:2019}'s
extensive taxonomy of bugs in deep learning systems
identified 
several causes of bugs that are specific to NN program,
including incorrect/incomplete models, 
wrong input data types, and training process issues.

\nicepar{Bugs in NN models.}
As we recalled in \autoref{sec:introduction},
a NN program implements in code a NN \emph{model}
that is trained on some \emph{data}, both of which 
can also be plagued by mistakes.
Hence, traditional software engineering approaches
to test generation~\cite{Sun:2019}, 
mutation testing~\cite{Hu:2019,Shen:2018}, 
fault localization~\cite{Eniser:2019}, 
and even automated program repair~\cite{Sohn:2019} 
have been applied to NN models and training data
to assess and improve their quality, robustness, and correctness.
Under this paradigm, 
bugs are revealed by \emph{adversarial examples},
e.g., two slightly different inputs that appear identical to the human eye but result in widely different classification by a trained model~\cite{Sun:2019}. 
Adversarial examples correspond to failing tests;
and fault localization and fixing
correspond to finding~\cite{Eniser:2019} and changing~\cite{Sohn:2019}
neuron weights in a model,
This kind of research is complementary to
our work on \annotest,
which is specific to NN programs
but focuses on testing
and finding faults in their code implementations.

\nicepar{Test oracles.}
Testing a program comprises three main steps~\cite{Ammann:2017}.
First, selecting concrete \emph{inputs} (arguments and pre-state);
second, \emph{executing} the program under test on those inputs;
and third, checking
whether the program behaved as expected while executing---in particular, whether its \emph{output}
(return values and post-state) is as expected.
The present paper's contribution, as well as the related work we discussed in the rest of this section,
concerns the first step: test-input generation.
In contrast, addressing the third step requires an \emph{oracle}: a mechanism to check the outcome of test execution;
thus, the problem of designing such mechanisms is known as the test oracle problem~\cite{Shahamiri:2009,Barr:2015,Oliveira:2014,Pezze:2014}.

Similarly as for test-input generation,
a key research challenge is \emph{automating} the generation of suitable oracles,
so as to reduce the required developer effort.
The simplest kind of test oracle are \emph{implicit oracles},
such as the crashing oracles we used for our experiments with \annotest.
More expressive automated test oracles may be derived from some kind of formal specification~\cite{Aichernig:1999},
such as assertions~\cite{Coppit:2005} and contracts~\cite{Araujo:2011},
as well from informal or semi-formal documentation written in natural language~\cite{Schwitter:2002,TranslatingCodeComments-ISSTA18}.
In absence of specifications, a practical option is building \emph{regression oracles}~\cite{Yoo:2012},
which check whether a new version of a program retains the same input/output behavior on the test inputs
as a previous version~\cite{Xie:2006};
test-input generators---like the aforementioned Randoop, EvoSuite, and Pynguin---%
are usually capable of building some kind of regression oracles automatically.

\section{Conclusions and Future Work}

The paper presented the \annotest approach
to generate
inputs that test NN programs
written in Python.
\annotest relies on code annotations that
precisely and succinctly describe the range of valid
inputs for the functions under test.
Using this information, \annotest can generate
tests that avoid spurious failures, and thus have a good
chance of exposing actual bugs.
In 
an experimental evaluation targeting 19 open-source NN programs,
\annotest
was able to reveal 94 bugs (including 63 previously known ones)
with an overhead of 6 annotations per tested function on average.

\nicepar{Future work.}
A natural continuation of the work on \annotest is 
extending \an to support more kinds of constraints.
As discussed in~\autoref{sec:rq3-answer},
most of the generator functions we wrote for our experiments
generate complex NN model objects such as tensors;
being able to specify such objects concisely would
further increase the applicability and convenience of using \annotest.

This paper's contributions address the test-input generation problem,
which is largely independent of the test-oracle problem
(see \autoref{sec:oracle_problem} and \autoref{sec:related-work}).
In future work, we may extend \annotest to add support for other kinds of oracles.
Since \annotest is based on annotations%
---a form of lightweight formal specification---%
adding \emph{postconditions} would be a natural way to do so.
Unlike the annotations currently supported by \annotest,
which act as constraints on the pre-state and hence require a matching generation mechanism,
postconditions are evaluated on a test's post-state, and hence can simply be evaluated to determine
whether the test is passing or failing.
Regression oracles are another kind of oracles that are commonly supported by
test generation tools such as Pynguin~\cite{Lukasczyk:2020};
\annotest could add support for a similar mechanism to generate \emph{regression tests},
whose assertion capture the post-state of the program under test,
and can be re-run on future versions of the program to determine whether its expected behavior has changed.
Given \annotest's focus, it could target regression oracles
that capture NN-specific properties~\cite{Zhang:2014,Ding:2017,Nejadgholi:2019}.

\section*{Acknowledgments}
Work partially supported by SNF grant 200021-182060 (Hi-Fi).

\bibliographystyle{elsarticle-num}

\printendnotes[itemize]

\end{document}